\documentclass[a4paper,11pt,preprintnumbers]{article}

\usepackage{jhep}

\usepackage[T1]{fontenc} % if needed
\usepackage{enumitem}
\usepackage[normalem]{ulem} % for \sout
\usepackage[capitalize]{cleveref}

\usepackage{journals}

\newcommand{\Bc}{B_{\star}}
\newcommand{\Ms}{M_{\star}}
\newcommand{\Rs}{R_{\star}}

\newcommand{\vp}{\bold{p}}
\newcommand{\ab}{\alpha\beta}
\newcommand{\tU}{\tilde{U}}
\newcommand{\diag}{\text{diag}}

\newcommand{\modify}[1]{\textcolor{black}{#1}}

\preprint{USTC-ICTS/PCFT-23-42}
\title{Magnetar-powered Neutrinos and Magnetic Moment Signatures at IceCube}

\author[a,b]{Vedran Brdar,}
\author[c]{Ting Cheng,}
\author[d]{Hao-Jui Kuan,}
\author[e,f]{Ying-Ying Li}

\affiliation[a]{Department of Physics, Oklahoma State University, Stillwater, OK, 74078, USA}
\affiliation[b]{CERN, Theoretical Physics Department, 1211 Geneva 23, Switzerland}
\affiliation[c]{Max-Planck-Institut f\"ur Kernphysik, Saupfercheckweg 1, 69117 Heidelberg, Germany}
\affiliation[d]{Max Planck Institute for Gravitational Physics (Albert Einstein Institute), 14476 Potsdam, Germany}
\affiliation[e]{Interdisciplinary Center for Theoretical Study, University of Science and Technology of China, Hefei, Anhui 230026, China}
\affiliation[f]{Peng Huanwu Center for Fundamental Theory, Hefei, Anhui 230026, China}

\emailAdd{vedran.brdar@okstate.edu}
\emailAdd{ting.cheng@mpi-hd.mpg.de}
\emailAdd{hao-jui.kuan@aei.mpg.de}
\emailAdd{yingyingli@ustc.edu.cn}

\abstract{The IceCube collaboration pioneered the detection of $\mathcal{O}{(\text{PeV})}$ neutrino events and the identification of astrophysical sources of high-energy neutrinos. 
In this study, we explore scenarios in which high-energy neutrinos are produced in the vicinity of astrophysical objects with strong magnetic field, such as magnetars. While propagating through such magnetic field, neutrinos experience spin precession induced by their magnetic moments, and this impacts their helicity and flavor composition at Earth.
Considering both flavor composition of high-energy neutrinos and Glashow resonance events we find that detectable signatures may arise at neutrino telescopes, such as IceCube, for presently unconstrained neutrino magnetic moments in the range between $\mathcal{O}(10^{-15})~\mu_B$ and $\mathcal{O}(10^{-12})~\mu_B$.}

\arxivnumber{2312.14113}

\begin{document}
\maketitle
\flushbottom

%=============================================================================
\section{Introduction}
\label{sec:intro}
%=============================================================================

Standard Model (SM) is an incredibly successful theory which nevertheless comes short
in several aspects, one of which is the explanation of non-vanishing neutrino masses. 
Massive neutrinos have non-vanishing magnetic moments ($\nu$MM) which are, for $m_\nu\sim$ eV, in the ballpark of $\mu\sim 10^{-20} \mu_B$ \cite{PhysRevLett.45.963,PhysRevD.16.1444,Petcov:1976ff,Zatsepin:1978iy,PhysRevD.25.766}. While such values are essentially experimentally unreachable \cite{Giunti:2014ixa}, there are several realizations beyond the Standard Model (BSM) which feature generation of large and testable $\nu$MM as well as the small neutrino masses \cite{SHROCK1982359,Lindner:2017uvt,Xu:2019dxe,Babu:2020ivd,Brdar:2020quo}. Large $\nu$MM can source the excess of neutrino scattering events \cite{Babu:2020ivd}, enhance neutrino decay rates \cite{Huang:2018nxj,Brdar:2023tmi}, impact the physics of the early Universe \cite{Li:2022dkc} and influence neutrino propagation in the Sun and supernovae \cite{PhysRevLett.45.963,Akhmedov:1987nc,PhysRevD.37.1368,Akhmedov:2022txm,Jana:2022tsa,Adhikary:2022phm,Sasaki:2023sza,Wang:2023nhh}. For latter, magnetic field plays a role in the transition between left-handed and right-handed neutrino states.

In the context of high-energy neutrinos, previous studies of $\nu$MM were chiefly focused on neutrino propagation through the intergalactic medium \cite{Kurashvili:2017zab,Alok:2022pdn,Lichkunov:2022mjf,Kopp:2022cug}. There, the idea is that despite the small magnetic field involved, neutrinos carrying large $\nu$MM can undergo efficient flavor conversion across large propagation distances. While the results of these analyses imply potentially observable effects, it should still be noted that the intergalactic magnetic fields presently come with rather large uncertainties. An alternative to studying the appearance of $\nu$MM effects over the cosmic distances is to investigate potential impact across much smaller scales, namely in the vicinity of neutrino production.
Clearly, this would require involvement of magnetic fields that are orders of magnitude larger compared to the intergalactic ones. There are several types of astrophysical objects that source such magnetic fields \cite{Turolla:2015mwa,Reimers:1995ia} and, if they also produce neutrinos, $\nu$MM effects may manifest. This was sketched in \cite{SinghChundawat:2022mll} for Dirac neutrinos passing through white dwarfs. 

In fact, astrophysical environment with large magnetic field is a prerequisite for the acceleration of charged particles which source high-energy neutrinos \cite{rach98,Bustamante:2020bxp,Hummer:2010ai}. In this work, we assume Majorana nature of neutrinos and adopt magnetars \cite{Kaspi:2017fwg} as a case study. We are motivated by the fact that $(i)$ magnetars source extremely strong magnetic fields and $(ii)$ it has been demonstrated in the literature that such objects produce high-energy neutrinos 
\cite{Dey:2017kly,Dey:2016psn,Zhang:2002xv,Murase:2009pg,carp20}. Therefore, considering high-energy neutrinos with large $\nu$MM in such astrophysical environments appears particularly motivating. We focus on two key phenomenological features at neutrino telescopes -- the flavor composition of high-energy neutrinos with energy $E \gtrsim 100$ TeV and the Glashow resonance events.  IceCube already pioneered in measuring both \cite{IceCube:2015gsk,HESEflavor,IceCube:2021rpz}.

The paper is organized as follows. In \cref{sec:theory}, we setup the framework for the flavor evolution of neutrinos carrying $\nu$MM in the magnetic field. In \cref{sec:astro} we focus on the magnetic field and its roles in neutrino production as well as estimate the impact of $\nu$MM in realistic astrophysical scenarios of interest. 
In \cref{sec.mag}, focusing on magnetar systems, we start by specifying representative magnetic field profile used in the analysis. Subsequently, we discuss details of our numerical simulations and show respective results. In \cref{sec:results} we utilize these results in order to assess $\nu$MM prospects at neutrino telescopes, namely its impact both to the flavor composition of high-energy neutrinos and the number of Glashow resonance events. We summarize in \cref{sec:conclusion}. Unless stated otherwise, all quantities are given in units $G=\hbar=c=1$. Indices $a$, $b$ denote spacetime components, $\alpha$, $\beta$ denote neutrino flavors and $j$, $k$ denote neutrino mass eigenstates.

%%%%%%%%%%%%%%%%%%%%%%%%%%%%%%%%%%%%%%%%%%%%%%%%%%%%%%%%%%%%%%%%%%%%%%%%%%%%%%%%
%=============================================================================
\section{The Formalism}
\label{sec:theory}
%=============================================================================
We base this study on the following effective interaction term 
\begin{align}
	\mathcal{L} \supset \mu_{\ab}\nu_{L,\alpha}\sigma_{ab}\overline{\nu}^c_{L,\beta}F^{ab}
	+ \rm h.c.~,
\label{NMM_operator}
\end{align} 
where $\mu_{\ab}$ is the component of the $\nu$MM matrix $\mu$, $\nu_{L}$ is the left-handed neutrino field, and $F^{ab}$ is the electromagnetic field strength tensor. 
In terms of density matrix $\rho(t,\vp)= |\nu(t,\vp)\rangle\langle \nu(t,\vp)|$, the dynamics of the neutrino state can be described by
\begin{align}\label{eq.EOM}
	\frac{d}{dt} \rho(t,\vp) = i\left[\mathcal{H},\rho(t,\vp)\right]+\mathcal{C}\,,
\end{align}
where the inelastic collision term $\mathcal{C}$ parametrizes neutrino flux attenuation, and the Hamiltonian $\mathcal{H}$ reads
\begin{align}\label{eq.Hami}
    \mathcal{H}= \mathcal{H}_{\rm vac}+\mathcal{H}_{\rm MSW}+\mathcal{H}_{\rm\nu MM}\,.
\end{align}
Here, $\mathcal{H}_{\rm vac}$, $\mathcal{H}_{\rm MSW}$ and $\mathcal{H}_{\rm \nu MM}$ quantify neutrino oscillations in vacuum, the Mikheyev-Smirnov-Wolfenstein (MSW) matter terms \cite{Wolfenstein:1977ue,Mikheyev:1985zog,Mikheev:1986wj}, and the $\nu$MM effects arising from \cref{NMM_operator}, respectively. On top of the flavor mixing induced by vacuum oscillations and MSW effects, additional helicity\footnote{For neutrinos with energies above $E \simeq 100$ TeV, that we will focus on in this work, helicity and chirality coincide for all practical purposes.} and flavor mixing will be induced by nonzero $\mu_{\alpha\beta}$ when neutrinos propagate through the background magnetic field. Hence, the basis state $|\nu(t,\vp)\rangle$ encodes the irreducible number of neutrino flavors and helicities. 

Notice that in building \cref{NMM_operator} we used only left-handed neutrino fields which already indicates that in this study we will be chiefly focused on Majorana neutrinos.
All the expressions below are valid for Majorana neutrinos; the Dirac neutrino case  follows straightforwardly and, for completeness, we outline it in \cref{sec:dirac}. For the Majorana case, we have three flavors with $\{\alpha,\beta \}= \{e, \mu, \tau\}$ and two helicities denoted as $\{h,h'\}= \{1 \,(\text{for } \nu_L), 2\,(\text{for } \bar{\nu}_L)\}$. The basis state $|\nu(t,\vp)\rangle$ is therefore a dimension-6 vector and $\mathcal{H}$ can be expressed as a $6 \times 6$ matrix with the entries given by     
\begin{align}
	\mathcal{H}_{\ab}^{11/22}  &= \frac{1}{2\vp}\sum_{j}U_{\alpha j}m_j^2U^{\dagger}_{j\beta} 
	\pm \delta_{\ab}V_{\alpha}\,, &   
	\mathcal{H}_{\ab}^{12}&= \left( \mathcal{H}_{\beta\alpha}^{21} \right)^{*}
	= \mu_{\ab}\,B_\bot\,e^{i\phi}\,.
\label{eq.H_Maj1}
\end{align}

The left term in Eq.~\ref{eq.H_Maj1} is responsible for standard neutrino oscillations. Here,  $m_j$ is the neutrino mass corresponding to the mass eigenstate $j$, $U$ is the leptonic mixing matrix, and $V$ is the MSW matter potential (minus sign for antineutrinos) which is diagonal in the flavor basis. The right term in Eq.~\ref{eq.H_Maj1} is the helicity-flipping (HF) term induced by $\nu$MM, where $B_\bot$ is the strength of the magnetic field projected to the plane perpendicular to the propagation direction of neutrinos, and the angle $\phi$ parameterizes the orientation of $B_\bot$ within that plane. We note that $\mu_{\ab}$ is asymmetric due to CPT symmetry (diagonal terms vanish) indicating that a neutrino with a flavor $\alpha$ will be converted to an antineutrino with a different flavor $\beta\neq\alpha$. The neutrino flavor evolution can only be performed numerically due to the complexity of $\mathcal{H}$. Nevertheless, different components of $\mathcal{H}$  may dominate at different characteristic length scales which can simplify the overall treatment, see discussions below.

\subsection{Characteristic Length Scales}
\label{subsec.charlength}
The length scales associated to the terms in \cref{eq.Hami} are given by $L_{jk}=2\pi/|E_j-E_k|$, where $E_j$ and $E_k$ denote the eigenvalues of the corresponding Hamiltonian component. The minimal length scale \modify{($L_{\rm ch}$)} is then determined by $\max\{E_j-E_k\}$. A crucial feature of having a HF term is that the dimension of $\mathcal{H}_{\rm\nu MM}$ is such that more combinations of $E_j$ and $E_k$ are possible in comparison to the standard scenario with $\mathcal{H}_{\rm vac}$ and $\mathcal{H}_{\rm MSW}$ terms; hence, multiple length scales may arise. In the absence of $\nu$MM effect, the dimension would be reduced to standard three flavors. For neutrinos traveling as separated mass eigenstates, the dimension may reduce for Dirac neutrinos if the diagonal entries $\mu_{\alpha\alpha}$ dominate. In such cases, the induced $\nu$MM effects would not be subject to mass-splitting suppression. However, for the Majorana case, since non-vanishing $\mu_{\alpha\beta}$ terms mix different flavors, such reduction cannot be made. Majorana neutrinos also suffer a large suppression of $\nu$MM effects from the mass-splitting terms when traveling over cosmic distances through regions with a weak intergalactic magnetic field of $\mathcal{O}(\mu G)$ \cite{Kopp:2022cug}.

From the left term in \cref{eq.H_Maj1} we infer that the minimal length scale for vacuum oscillations is set for the largest $m_j^2-m_k^2$ (atmospheric mass squared difference)
\begin{align}\label{eq.length_vac}
    L_{\rm vac} = \frac{4\pi E}{\max \{\Delta m_{jk}^2\}} \simeq 9.9 \times 10^{8} \, {\rm{km}}
	\left(\frac{\Delta m^2_{32}}{2.4 \times 10^{-3} \rm{eV}^2}\right)^{-1} 
	\left(\frac{E}{1\, \rm{PeV}}\right)\,.
\end{align}
As far as the MSW term is concerned, the refractive length is given by $2\pi/|V_\alpha-V_\beta|$. In particular, $V_{\nu_e}=\sqrt{2}G_F(n_e-n_n/2)$ and $V_{\nu_\mu}= V_{\nu_\tau}= -\sqrt{2}/2G_Fn_n$, where $n_e$ ($n_n$) denotes number density of electrons (neutrons).
Due to the opposite signs for neutrinos and antineutrinos in \cref{eq.H_Maj1}, $V_\alpha \in \{V_{\nu_e},V_{\nu_\mu, \nu_\tau},V_{\bar{\nu}_e}\equiv -V_{\nu_e},V_{\bar{\nu}_\mu,  \bar{\nu}_\tau}\equiv-V_{\nu_\mu, \nu_\tau} \}$. The minimal length scale associated to the matter term reads
\begin{align}\label{eq.length_MSW_M}
    L_{\rm MSW}^{\rm M}&=\frac{\sqrt{2}\pi}{G_F \max \{n_e,n_n,|2n_e-n_n|\}} \\ \nonumber
    & \simeq 1.8 \times 10^{6} \, \text{km}
	\left(\frac{\max \{Y_e,1-Y_e,|1-3Y_e|\}}{0.9}\right)^{-1}
    \left(\frac{\rho_B}{0.01\,\text{g}\,\text{cm}^{-3}} \right)^{-1},
\end{align}
where $\rho_B$ is the density of baryon matter, and $Y_e$ is the fraction of electrons. Note the relation $n_e=Y_e n_b$, where $n_b=n_p+n_n$ is the baryon number density and $n_p = n_e$ is the proton number density. Here, the benchmark value taken for $\rho_B$ aligns with numerical simulations for magnetar born in a merger event \cite{Kawaguchi:2022bub}. Furthermore, as elaborated in the following, larger values of $\rho_B$ would lead to significant attenuation of the neutrino flux.

For neutrinos carrying $\mathcal{O}(\text{PeV})$ energy, the cross sections for neutrino scattering off nuclei, $A$, are of the order $10^{-33}$ cm$^2$ \cite{Bustamante:2017xuy,Formaggio:2012cpf} which is comparable to $G_F$; and the cross section for neutrino-electron scattering is smaller. In passing, let us stress that $\nu$MM contribution to neutrino scattering is negligible since it is proportional to $\mu_{\alpha\beta}^2 \ll G_F$. Therefore, the length scale of the MSW effect compared with the collision length, given by $L_{\rm \mathcal{C}}=1/(\sigma_{\nu A}n_A)$, can be estimated as 
\begin{equation}\label{eq.Lratio}
    \frac{L_{\rm MSW}^{\rm M}}{L_{\mathcal{C}}} \simeq \frac{\sqrt{2}\pi \sigma_{\nu A}}{G_F}\, 
    = \frac{\sigma_{\nu A}(E)}{6.4 \times 10^{-33} \text{cm}^2}\,,
\end{equation}
where in making this comparison we dropped $\mathcal{O}(1)$ factors depending on particle number densities entering $L_{\rm \mathcal{C}}$ and $L_{\rm MSW}^{\rm M}$. One can thus infer that, for $E\sim\mathcal{O}(\text{PeV})$, MSW effect becomes relevant at the length scale for which neutrino absorption also becomes significant. This is the rationale behind not considering MSW effects in this work, as we are not interested in the case where total neutrino fluxes get strongly attenuated. For instance, if neutrinos are produced $10^5$ km above the magnetar, the flux would be attenuated when $\rho_B > 0.01$ g~cm$^{-3}$ at the production site. 

\begin{figure}
    \centering
    \includegraphics[scale=0.5]{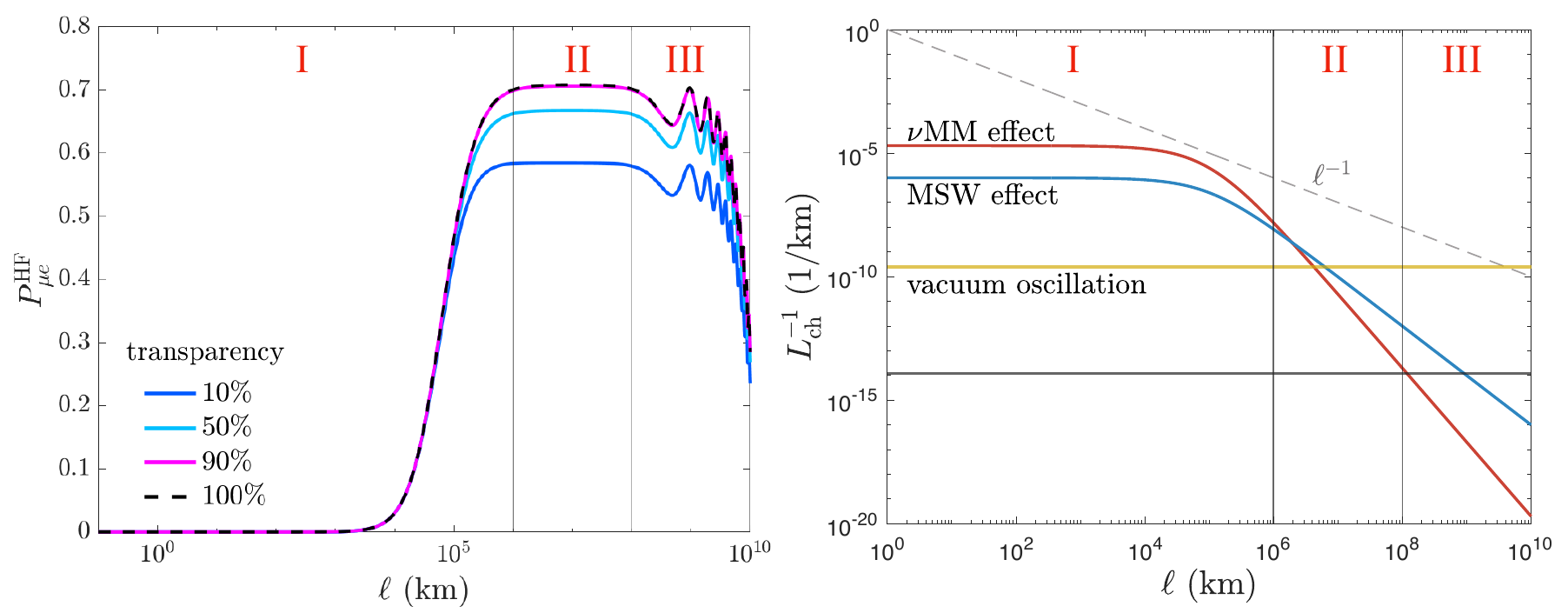}
    \caption{\modify{\emph{Left plot}: Transition probability $\nu_\mu \rightarrow \bar{\nu}_e$ obtained by numerically evolving the full Hamiltonian for 1 PeV neutrinos. \emph{Right plot}: The inverse characteristic length scales of the $\nu$MM effect ($L_{\rm \nu MM}^{\rm M,\,-1}$, red), MSW effect ($L_{\rm MSW}^{-1}$, blue), vacuum oscillation for 1 PeV neutrinos  ($L_{\rm vac}^{-1}$, yellow), and the inverse length itself (grey). The regions labeled as I, II and III correspond to the $\nu$MM-dominated, ``quiet'' and vacuum-dominated regions, respectively. In both plots, the magnetic field strength and the matter density are localized within $R=10^5$ km. For all lines, the $\nu$MM length scale is fixed, such that $\int d\ell \, L_{\rm \nu MM}^{\rm M,\,-1} = 1$. The length scale for MSW effect is taken for different values of $\rho_B$, such that the transparency ($1- \int d\ell \, L_{\mathcal{C}}^{-1}$) takes the respective labeled values in the left plot, and 90\% in the right plot.}}
    \label{fig.Evol}
\end{figure}

Obtaining eigenvalues of $\mathcal{H}_{\rm\nu MM}$ for Majorona neutrinos is non-trivial due to the asymmetricity of the $\nu$MM matrix. Nonetheless, we found  $\{E_j, E_k\}\in\{0,a_\nu B_\bot\}$, where degeneracy is implied and $a_\nu^2 = \mu_{e\mu}^2 + \mu_{e\tau}^2 + \mu_{\mu\tau}^2$, see \cref{sec.2.2}. Hence, the minimal length scale can be estimated as
\begin{align}\label{eq.length_NMM_M}
    L_{\rm \nu MM}^{\rm M}=\frac{2\pi}{a_\nu B_\bot} 
    \simeq 1.2 \times 10^{5} \, \text{km}
	\left(\frac{a_\nu}{10^{-12}\mu_B}\right)^{-1}
    \left(\frac{B_\bot}{10^6\, \text{G}} \right)^{-1}.
\end{align}

Given that $L_{\text{vac}}$ is dependent on neutrino energy but uniform over space, in contrast to $L_{\rm \nu MM}^{\rm M}$ that depends on the magnetic field strength, different effects may dominate in different regions. In this work, our goal is not to perform an exhaustive analysis of realistic magnetic field and matter density profiles that are extremely difficult to model. Instead, we discuss a scenario which captures the most important features of the relevant physical systems. In particular, we consider a dipolar magnetic field (see \cref{sec.mag}) for $r\leq R$, and $B_{\bot} \propto (r/R)^{-3}$ for $r > R$. Here, $r$ is the distance  from the magnetar as the neutrinos travel, and $R$ corresponds to the production site.

For systems with $R\ll L_{\rm vac}$, we find two typical distances, $L_1$ and $L_2$, such that $L_{\rm \nu MM}^{\rm M}\ll L_{\rm vac}$ for $l\leq L_1$, and $L_{\rm \nu MM}^{\rm M}\gg L_{\rm vac}$  for $\ell>L_2$. \modify{The existence of $L_1$ and $L_2$ indicates that $\nu$MM and vacuum effects can be decoupled. \cref{fig.Evol} shows the probability of a helicity flip from $\nu_\mu\rightarrow\bar{\nu}_e$, and a comparison of characteristic length scales, for a simple example where PeV neutrinos are produced at $r=R$ and then travel outward across distance $\ell = r - R$. Here, the matter density profile is taken as $\rho_B \propto (r/R)^{-2}$, $Y_e=1$ (conservatively for the matter effect), and $\sigma_{\nu A} = 6\times 10^{-33}$ cm$^2$. 
The propagation is split by $L_1$ and $L_2$ into three regions with $\nu$MM and vacuum effects dominating in region I and III, respectively. 
Note that in region II, although the three effects may be comparable with one another, none of them is large enough to induce a considerable flavor/helicity transition. In other words, since all length scales exceed the size of the region ($\min\{L_{\rm ch}\}>L_2-L_1$ within region II), none of them can amount to any significancy. We refer to this region as a ``quiet'' one.
By comparing results with various $\rho_B(r=R)$ in the left plot of \cref{fig.Evol}, one can also infer how the MSW effect would affect the flavor evolution. In the figure,  labels of 100\%, 90\%, 50\%, 10\%, correspond to $\rho_B(r=R)=0,\,0.02,\,0.1,\,0.9$ g/cm$^{-3}$, respectively. For all  cases, the evolution of neutrino transition probability over the region II is not inducing any significant effect.}

\modify{Owing to the decoupling between $\nu$MM and vacuum effects,} the neutrino transition probability, from production to detection at distance $L$, can thus be approximately written as
\begin{align} \label{eq.decouple1}
    P_{\ab}^{hh'}(0\rightarrow L) 
    \simeq \sum_{\gamma\delta h''}
    P_{\alpha\delta}^{hh''}(0\rightarrow L_1; \mathcal{H}_{\rm \nu MM})
    \times P_{\delta\gamma}^{h''h'}(L_1\rightarrow L_2)
    \times P_{\gamma\beta}^{h'h'}(L_2\rightarrow L; \mathcal{H}_{\rm vac})\,.
\end{align}
Furthermore, when $L_2-L_1\ll L_{\rm vac}$, this relation effectively simplifies to
\begin{align} \label{eq.decouple2}
    P_{\ab}^{hh'}(0\rightarrow L) 
    \simeq \sum_{\delta}
    P_{\alpha\delta}^{hh'}(0\rightarrow L_{\rm cut}; \mathcal{H}_{\rm \nu MM})
    \times P_{\delta\beta}^{h'h'}(L_{\rm cut}\rightarrow L; \mathcal{H}_{\rm vac})\,.
\end{align}
Here, $L_1 < L_{\rm cut} < L_2$ and we found that results only mildly depend on the choice of $L_{\rm cut}$. In our calculations, we employ \cref{eq.decouple2}.

%%%%%%%%%%%%%%%%%%%%%%%%%%%%%%%%%%%%%%%%%%%%%%%%%%%%%%%%%%%%%%%%%%%%%%%%%%%%%%%%%%%%%
\subsection{$\nu$MM  effect}
\label{sec.2.2}
Let us calculate the first term in Eq.~\eqref{eq.decouple2} for which $\nu$MM  effect dominates. The density matrix, introduced in \cref{eq.EOM}, will evolve from initial time $t_0$ by an infinitesimal time $\delta t$ as
\begin{equation}
	\rho(\delta t+t_0,\vp) = |\nu(\delta t+t_0,\vp)\rangle\langle \nu(\delta t+t_0,\vp)| = e^{-iA(\delta t)}  \rho(t_0,\vp) \, e^{iA(\delta t)}\,,
\end{equation}
where the exponent of the time evolution operator reads
\begin{align}
    A(\delta t) = 
    \begin{pmatrix}
        0 & \mu B_\bot e^{i\phi} \\
        \mu^\dagger B_\bot e^{-i\phi}  & 0
    \end{pmatrix} \delta t\,.
\end{align}
Focusing on Majorona neutrinos we define the normalized $\nu$MM tensor
\begin{align}
	\hat{\mu} = \frac{1}{|a_\nu|}\mu=\frac{1}{|a_\nu|} 
	\begin{pmatrix}
	0 & \mu_{e\mu} & \mu_{e\tau}\\ 	
	-\mu_{e\mu} & 0 & \mu_{\mu\tau}\\ 
	-\mu_{e\tau} & -\mu_{\mu\tau} & 0\\
	\end{pmatrix}\,,
\end{align}
for which it is convenient to introduce a unitary matrix $\tilde{U}$ (not to be confused with the leptonic mixing matrix $U$ aforementioned) such that  $\tU \, |\hat{\mu}|^2 \,\tU^{\dagger}=\diag(0,1,1)$. After some straightforward derivations, we find
\begin{align} \label{eq.time_opt_M}
    e^{-iA(\delta t)}=    
    \begin{pmatrix}
        \tU^{\dagger} \cos\hat{\theta} (\delta t)\,\tU
        & 
        i\tU^{\dagger}  \sin\hat{\theta}(\delta t) \,\tU \,\hat{\mu} e^{i\phi}
        \\     
        -i\hat{\mu}^\dagger e^{-i\phi} \tU^{\dagger}  \sin\hat{\theta} (\delta t) \,\tU 
        &
        \tU^{\dagger}  \cos\hat{\theta}(\delta t) \,\tU 
    \end{pmatrix}  \,,
\end{align}
with $\hat{\theta}(\delta t) = \diag(0,a_\nu B_\bot \delta t,a_\nu B_\bot \delta t)$.

For the scenario where $\phi$ is adiabatically evolving, namely when 
\begin{align}\label{eq.adiab}
    |\dot{\phi}/\phi|\ll a_\nu B_\bot,
\end{align}
we can write down the density matrix at time $t$ as $\rho(t,\vp) = e^{-iA(t)}  \rho(0,\vp) \, e^{iA(t)}$. Furthermore, within the $\nu$MM length scale ($t<L_{\rm \nu MM}$), the time evolution operator reads
\begin{align}\label{eq.evo.oper}
    e^{-iA(t)}\simeq    
    \begin{pmatrix}
        \tU^{\dagger} \cos\hat{\theta} (t)\,\tU
        & 
        i\tU^{\dagger}  \sin\hat{\theta}(t) \,\tU \,\hat{\mu} e^{i\phi}
        \\     
        -i\hat{\mu}^\dagger e^{-i\phi} \tU^{\dagger}  \sin\hat{\theta} (t) \,\tU 
        &
        \tU^{\dagger}  \cos\hat{\theta}(t) \,\tU 
    \end{pmatrix}\,,
\end{align}
with
\begin{align} \label{eq.theta_nu}
    &\hat{\theta}(t) = \diag(0,a_\nu,a_\nu) 
    \int_0^{t} \, dt' \, B_\bot(t') \equiv (0, \theta_{\nu}, \theta_{\nu} )\,.
\end{align}
The transition probability from $\nu_\alpha^{h}$ to $\nu_\beta^{h'}$ can then be calculated using
\begin{align}
 P_{\ab}^{hh'} = {\rm{Tr}}\left[\langle\nu_\beta^{h'}|\rho^{(\alpha,h)}(t)|\nu_\beta^{h'}\rangle\right]\,.
\end{align}
Given the above, for the helicity-conserving (HC) case with $h=h'$ we find 
\begin{equation} \label{eq.HC_FTP}
    P_{\ab}^{\rm HC} = {\rm{Tr}}\left[\big|\langle\nu_\beta|\tU \cos(\hat{\theta})\tU^{\dagger}|\nu_\alpha\rangle\big|^2\right]
    = \sum_{j,k} \tU_{\alpha j}^* \tU_{\beta j}\tU_{\alpha k} \tU_{\beta k}^* \Phi_{jk} \cos\hat{\theta}_j\cos\hat{\theta}_k\,,
\end{equation}
while for HF with $h\neq h'$ the transition probability reads
\begin{equation}\label{eq.HF_FTP}
    P_{\ab}^{\rm HF} = {\rm{Tr}}\left[\big|\langle\nu_\beta|\tU \sin(\hat{\theta})\tU^{\dagger}\hat{\mu}_\nu|\nu_\alpha\rangle\big|^2\right]
    = \sum_{j,k,\gamma,\delta} \hat{\mu}_{\alpha \gamma}^* \hat{\mu}_{\alpha \delta} \tU_{\gamma j}^* \tU_{\beta j}\tU_{\delta k}  \tU_{\beta k}^* \Phi_{jk} \sin\hat{\theta}_j\sin\hat{\theta}_k\,.
\end{equation}
Here, $0\leq\Phi_{jk}\leq 1$ and $\Phi_{jj}=1$, are the decoherence terms describing the overlap between $j$ and $k$ eigenstates in the phase space \cite{Cheng:2022lys}. 
In contrast to the case of vacuum oscillations, the time evolution in $\nu$MM scenario cannot be diagonalized with real eigenvalues, and is instead described through the rotation matrices $\tU$. This implies that there will always be oscillation effect from the $\Phi_{jj}$ term even if $\Phi_{jk}=0$ for $k\neq j$. Furthermore, the fact that $\nu$MM of Majorona neutrinos possesses an additional asymmetricity further mitigates decoherence effects. In particular, since one of the $\nu$MM eigenvalues is zero and the other two are degenerate, only the terms corresponding to the same mass eigenstates ($\Phi_{jk}=1$) will survive after inserting Eq.~\eqref{eq.theta_nu} into Eq.~\eqref{eq.HF_FTP}. Hence, unlike for the vacuum oscillations where decoherence arises due to the separation of wave packets for different mass eigenstates ($\Phi_{jk} < 1$), there will be no such decoherence effects entering in the Majorana neutrino HF transition probability. 

\begin{figure}
	\centering
	\includegraphics[scale=0.45]{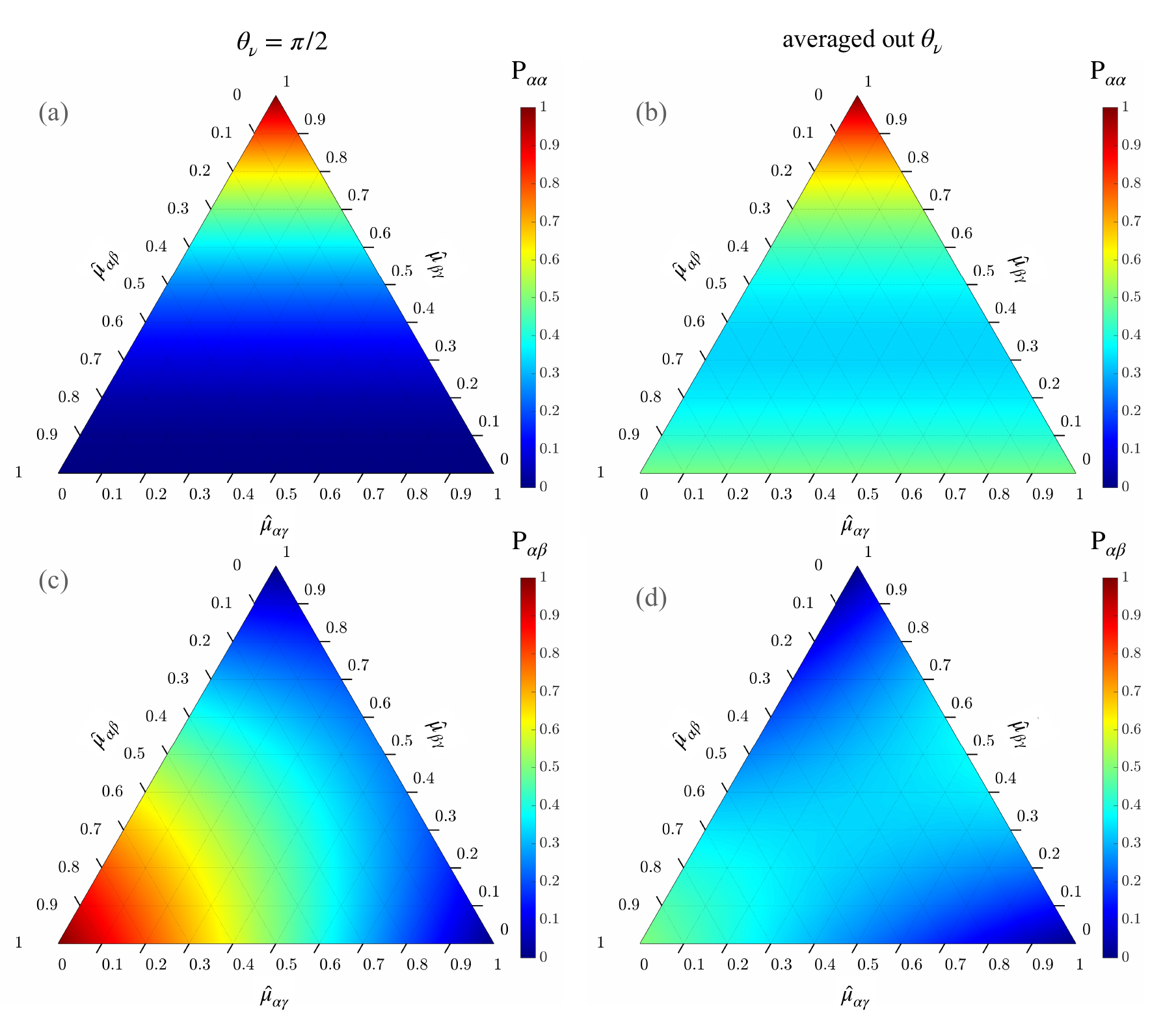}
	\caption{The ternary diagrams show how the structure of the $\nu$MM matrix with real values (relative magnitudes of the entries shown on the axes) impacts the oscillation probabilities. The panels $(a)$ and $(b)$ show the survival, while the panels $(c)$ and $(d)$ show flavor transition probability. In the left panels we assumed that the HF effect is maximized (i.e. $\theta_\nu=\pi/2$) while the right panels correspond to the case where $\theta_\nu$ is averaged out.}
    \label{fig.FlavourStructure}
\end{figure}

Let us now investigate HC and HF transition probabilities for $\theta_\nu \ll 1$. In that case,  the two-step transitions, e.g. $\nu_e \rightarrow \bar{\nu}_\mu \rightarrow \nu_\tau$, can be neglected so the flavor transitions for the HC case should vanish, unlike in the HF case. This can be checked by starting from \cref{eq.HC_FTP,eq.HF_FTP}
and plugging in $\cos\hat{\theta}_j \simeq 1$ and $\sin\hat{\theta}_j \simeq \hat{\theta}_j$ for all $j$. We find
\begin{align}
    &P_{\ab}^{\rm HC} 
    \simeq \sum_{j,k} \tU_{\alpha j}^* \tU_{\beta j}\tU_{\alpha k} \tU_{\beta k}^*=  \delta_{\ab}\,, \notag\\
    &P_{\ab}^{\rm HF}  
    \simeq \theta_\nu^2 \sum_{j,k,\gamma,\delta} \hat{\mu}_{\alpha \gamma}^* \hat{\mu}_{\alpha \delta} \tU_{\gamma j}^* \tU_{\beta j}\tU_{\delta k}  \tU_{\beta k}^* 
    =  \theta_\nu^2 \, |\hat{\mu}_{\ab}|^2 \,.
   \label{eq:small}
\end{align}
The optimal HF scenario ($\theta_\nu=\pi/2$), is shown in the left panels of Fig.~\ref{fig.FlavourStructure} for various flavor structures of $\nu$MM. Similarly, in the right panels we show the case where equilibrium is reached ($\theta_\nu$ is averaged out).  For both cases, the upper panels show flavor-conserving probability $P_{\alpha\alpha}=P_{\alpha\alpha}^{\rm HC}+ P_{\alpha\alpha}^{\rm HF}$ and the lower ones corresponds to flavor-changing probability $P_{\ab}=P_{\ab}^{\rm HC}+ P_{\ab}^{\rm HF}$ ($\alpha \neq \beta$). 

With only one non-vanishing term $\hat{\mu}_{\beta\gamma}$, the transition occurs among flavors $\beta$ and $\gamma$ leaving the $\nu_\alpha$ state unaffected; this is evident from the top corner in the left panels. Namely, $P_{\alpha\alpha}=1$ while $P_{\ab}=P_{\alpha\gamma} = 0$ for $\alpha \neq \beta \neq \gamma$. In the lower left corner of panels $(a)$ and $(c)$, $\hat{\mu}_{\ab} = 1$ yields a 100$\%$ transition from $\nu_\alpha$ to $\nu_\beta$ in the optimal $\theta_\nu = \pi/2$ scenario. In contrast, in the scenario where $\theta_\nu$ is averaged out, $\nu_\alpha$ and $\nu_\beta$ would each have an equal share; see the lower left corner in the right panels. In fact, such an equilibrium feature holds in general; for instance, the center of panels $(b)$ and $(d)$ indicates that all flavors have a $1/3$ share when $\hat{\mu}_{e\mu} = \hat{\mu}_{e\tau} = \hat{\mu}_{\mu\tau}$. In general, it can be shown that the oscillation probabilities are unitary
\begin{equation}
    \sum_{\beta} P_{\ab}^{\rm HC}    = \sum_{j} |\tU_{\alpha j}|^2 \cos^2\hat{\theta}_j \,\,\,,\hspace{0.4cm} \sum_{\beta} P_{\ab}^{\rm HF}    = \sum_{j} |\tU_{\alpha j}|^2 \sin^2\hat{\theta}_j
    \Longrightarrow \sum_{\beta} P_{\ab}^{\rm HC} + P_{\ab}^{\rm HF} = 1\,,
\end{equation}
as also evident from panels $(a)$ and $(c)$, and panels $(b)$ and $(d)$, respectively.

%%%%%%%%%%%%%%%%%%%%%%%%%%%%%%%%%%%%%%%%%%%%%%%%%%%%%%%%%%%%%%%%%%%%%%%%%%%%%%%%%

%%%%%%%%%%%%%%%%%%%%%%%%%%%%%%%%%%%%%%%%%%%%%%%%%%%%%%%%%%%%%%%%%%%%%%%%%%%%%%%%

%=============================================================================
\section{Roles of the magnetic field}
\label{sec:astro}
%=============================================================================
In astrophysical environments, the magnetic field makes an impact to the energy and flavor composition of produced neutrinos in a rather complicated manner. In particular, possible progenitors of high-energy neutrinos are relativistic protons and the magnetic field plays the dominant role in their acceleration (\cref{sec.3.1}). Therefore, the energy budget inherited by neutrinos will be limited by the magnetic field strength. The amount of energy that will be transferred to  neutrinos depends also on the sequential production processes: proton collisions first generate mesons which then decay to neutrinos. During these steps, the magnetic field works to dissipate the energy from charged particles mainly via synchrotron radiation (\cref{sec.3.2}). While the aforementioned aspects are related to the energy of neutrinos, note that the strength of the $\nu$MM effect is also closely associated to the magnitude and spread of the magnetic field, as was elaborated in \cref{sec.2.2}. In this section, we discuss multifaceted roles of magnetic field, and based on the parameters on the so called Hillas plot \cite{hill84,Ptitsyna:2008zs}, which shows the size of the accelerator region and magnetic field strength for various astrophysical environments, we will estimate (\cref{sec.3.3}) realistic strength of $\nu$MM effects.

\subsection{Acceleration of protons}
\label{sec.3.1}
Relativistic protons have been long thought as triggers of processes producing high-energy neutrinos carrying $\mathcal{O}(\text{PeV})$ energy through, e.g. proton-photon ($p\gamma$) and/or proton-proton ($pp$) collisions with the subsequent meson decays. The protons can be accelerated by magnetic field and can in principle acquire energy up to \cite{hill84,rach98}
\begin{align}\label{eq.Hillas}
    E_{\rm max}^p &= \eta eBR\Gamma
    \simeq 10^5
    \left( \frac{\eta}{0.1} \right)
    \left( \frac{B}{3\times 10^5 {\rm G}} \right)
    \left( \frac{R}{10^5 {\rm km}} \right)
    \left( \frac{\Gamma}{10^3} \right)
    \,\,{\rm PeV}\,.
\end{align}
Here, $R$ is the size of the acceleration region in the inertial frame of reference, $B$ is the strength of magnetic field that proton experiences when it reaches the energy of $E^p_{\rm max}$, $\Gamma$ is the Lorentz factor and $\eta$ is the acceleration efficiency (lower for shorter acceleration times) tied to the acceleration mechanism and the velocity of the moving source. A portion of the proton's energy can be handed over to the generated neutrino; in average, the attainable neutrino energy can be estimated as $E_{\rm max}^p/20$ \cite{Bustamante:2020bxp,Hummer:2010ai,Hummer:2011ms,Zhang:2012qy}.

%%%%%%%%%%%%%%%%%%%%%%%%%%%%%%%%%%%%%%%%%%%%%%%%%%%%%%%%%%%%%%%%%%%%%%%%%%%%%%
\begin{figure}[t!]
    \centering
	\includegraphics[width=.67\columnwidth]{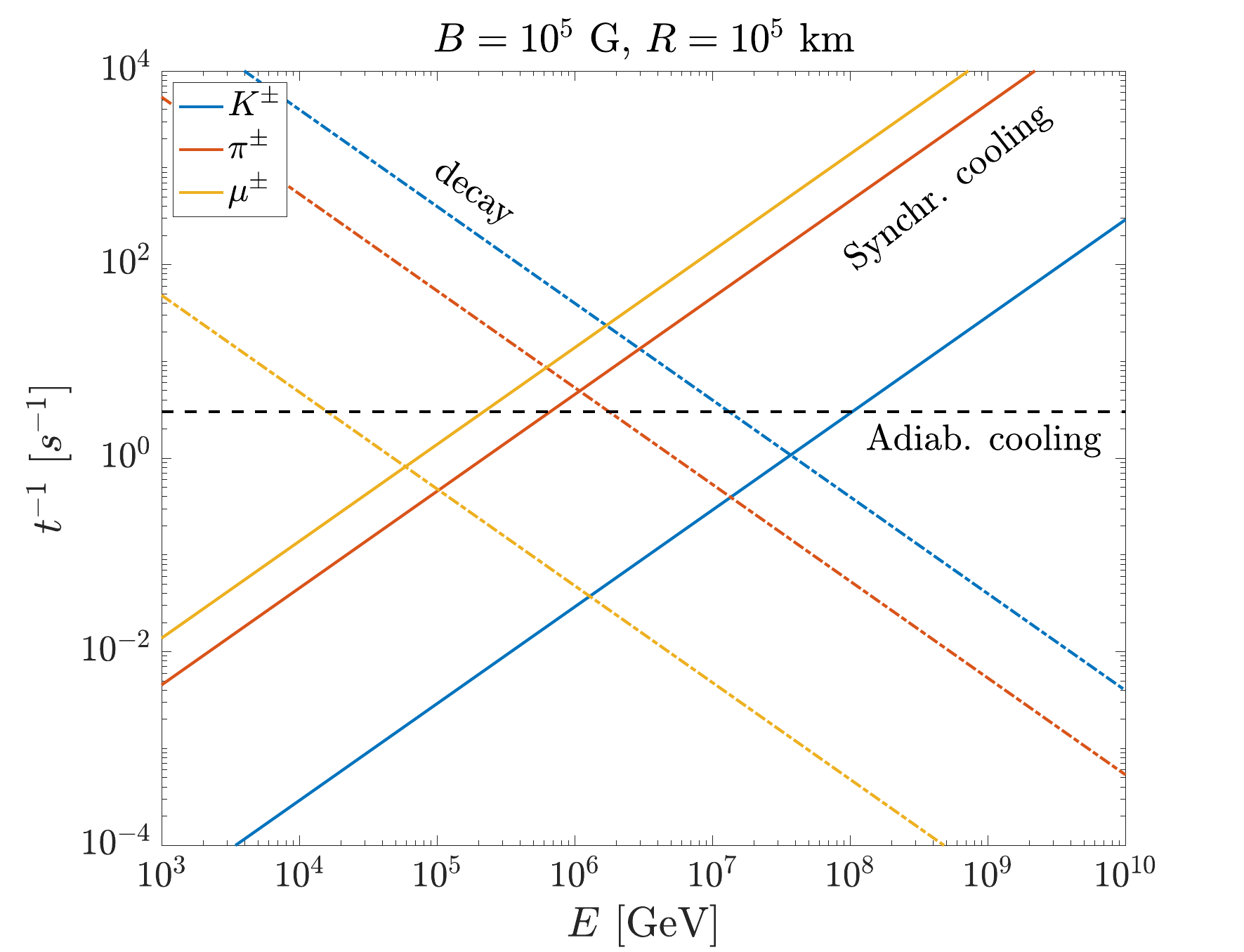}
	\caption{Adiabatic (black dashed) and synchrotron (solid) cooling timescales for secondary mesons, namely $\mu^{\pm}$, $\pi^{\pm}$, and $K^{\pm}$. Their decay timescales are also shown (dash-dotted). These results are obtained following \cite{Hummer:2010ai}, with $B=10^{5}$~G at $R=10^5$~km. }
    \label{fig.cooling}
\end{figure}
%%%%%%%%%%%%%%%%%%%%%%%%%%%%%%%%%%%%%%%%%%%%%%%%%%%%%%%%%%%%%%%%%%%%%%%%%%%%%%%

\subsection{Cooling}
\label{sec.3.2}
The condition in \cref{eq.Hillas} does not take into account various cooling processes experienced by protons and mesons. Cooling effects do not only affect the energy of neutrinos, but also alter the neutrino flavor composition since some decay processes may become irrelevant in producing neutrinos of very high energy. The timescale of cooling or dissipation, $t_{\rm dis}$, is generally jointly set by the two dominant energy loss mechanisms: the adiabatic loss due to the expansion of the shell (with the timescale $t_{\rm ad}$) and the synchrotron loss of the protons (with the timescale $t_{\rm syn}$) through
\begin{align}
    t_{\rm dis}^{-1}=t_{\rm syn}^{-1}+t_{\rm ad}^{-1}\,.
\end{align}
Depending on the acceleration ($t_{\rm acc, p}$) and cooling ($t_{\rm dis,p}$) timescales of protons, and the cooling ($t_{\rm dis,m}$) and decay ($t_{\rm dec,m}$) timescales of mesons, there are in principle three options for the generation of high-energy neutrinos:
\begin{enumerate}[label=$(\roman*)$]
    \item  For $t_{\rm acc, p}>t_{\rm dis,p}$, the protons are unable to be sufficiently accelerated, thus no high-energy neutrinos are expected.
    \item For $t_{\rm acc, p}<t_{\rm dis,p}$, and $t_{\rm dec,m}>t_{\rm dis,m}$, protons can be highly accelerated and subsequently produce energetic mesons through $pp$ and/or $p\gamma$ collisions, but mesons will not have enough time to pass over their energy to daughter particles before cooling down.
    \item For $t_{\rm acc, p}<t_{\rm dis,p}$, and $t_{\rm dec,m}<t_{\rm dis,m}$, protons can yield energetic mesons, and the latter decay to high-energy neutrinos. Channels that fall in this category may determine the initial flavor composition. For magnetar-powered gamma-ray burst (GRB), this usually happens $10^4$--$10^5$~km away from the magnetar, where the magnetic field has reduced from $\mathcal{O}(10^{16})$ G around the star's surface to $\mathcal{O}(10^{5})$ G \cite{Kumar:2014upa}.

\end{enumerate}
For $(ii)$ and $(iii)$, the comparison between meson decay and cooling timescales is shown in Fig.~\ref{fig.cooling} (c.f.~\cite{Hummer:2010ai,Winter:2011jr}). We see that muons with energy larger than $\mathcal{O}(10)$ TeV will be damped by synchrotron radiation before they decay, while pions with energy up to $\sim1$\,PeV can decay before losing significant energy via cooling. Due to their higher decay rate and larger mass, kaons with energy up to $50$ PeV can decay before experiencing significant energy loss. Therefore, since the cooling process is non-trivial and energy dependent, we will discuss all typical possibilities in \cref{sec:results}.

\subsection{Estimated magnitude of $\nu$MM effect}\label{sec.3.3}
%%%%%%%%%%%%%%%%%%%%%%%%%%%%%%%%%%%%%%%%%%%%%%%%%%%%%%%%%%%%%%%%%%%%%%%%%%%%%%%%%%%%%%%
\begin{figure}[t!]
    \centering
	\includegraphics[width=.7\columnwidth]{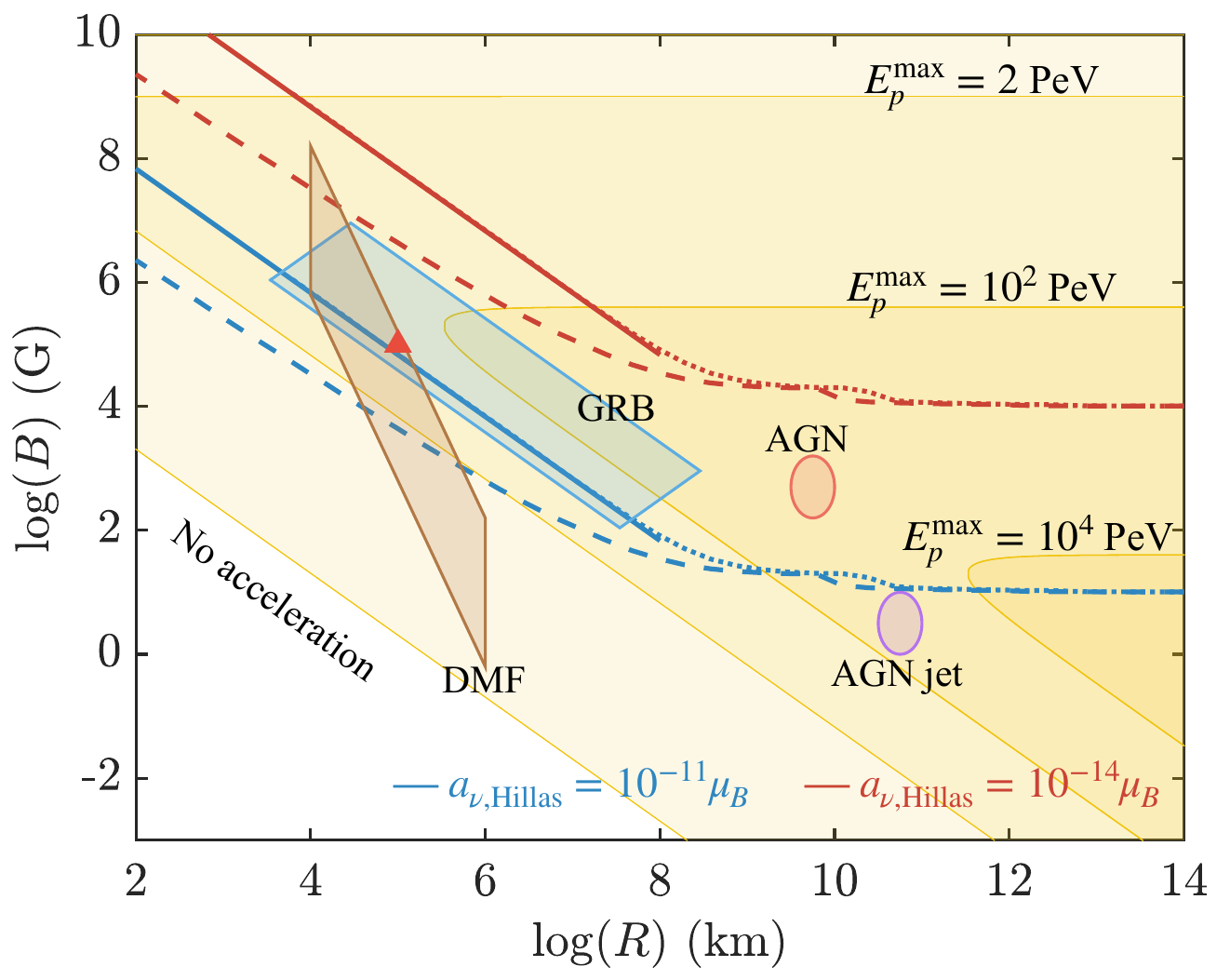}
	\caption{The Hillas plot. We show lines corresponding to $a_{\nu, \rm Hillas}=10^{-11} \mu_B$ \modify{(blue)} and $a_{\nu, \rm Hillas}=10^{-14} \mu_B$ \modify{(red)}. \modify{The dashed (dotted) lines show numeric results for $\alpha=1 (3)$, for neutrinos with 1 PeV energy. The solid lines stem from \cref{eq.Hillas_NMM} for the region where \cref{eq.decouple2} is applicable. The yellow contour lines show the maximum proton energies which are able to generate neutrino energy above 100 TeV, considering cooling effects and an acceleration efficiency of $\eta=0.1$.} The \modify{brown} region features dipolar magnetic field \cite{Mastrano:2013jaa,Petri:2015oaa} and the blue region represents GRB case \cite{pira99,Murase:2005hy}. The red triangle is the considered benchmark.}
    \label{fig.Hillas}
\end{figure}
%%%%%%%%%%%%%%%%%%%%%%%%%%%%%%%%%%%%%%%%%%%%%%%%%%%%%%%%%%%%%%%%%%%%%%%%%%%%%%%%%%%%%%%
Based on \cref{eq.theta_nu}, if neutrinos experience a field proportional to $\ell^{-\alpha}$, where $\alpha > 1$, then 
\begin{align}
\label{eq.thetanu_int}
    \theta_\nu =  a_\nu \, B \, \int_{R}^{L_{\rm cut}} dr \,
    \left (\frac{r}{R}\right)^{-\alpha}
    \simeq \frac{1}{\alpha - 1} \, a_\nu B R\,,
\end{align}
where $L_{\rm cut}\gg R$ fulfills the criteria given in \cref{eq.decouple2}.
This relation is determined through $B$ and $R$ and thus can be related to the maximal proton energy in \cref{eq.Hillas}. Using \cref{eq.thetanu_int} we can estimate $a_\nu$ for which we expect appreciable $\nu$MM effects. To this end, we set $\theta_\nu = 1$. Then, for the potential reach of $a_\nu$, denoted by $a_{\nu, \rm Hillas}$, we find 
\begin{equation} \label{eq.Hillas_NMM}
    a_{\nu, \rm Hillas} = 3.4 \times 10^{-12} \mu_B \, 
    \xi
    \left(\frac{ 10^5~\text{G}}{B}\right)
    \left( \frac{10^5~{\rm km}}{R} \right)\,,
\end{equation}
where we have defined $\xi= \alpha-1$. \modify{If the condition for decoupling of
$\nu$MM and vacuum effects is no longer met, the interplay with vacuum oscillation yields a more complicated form than \cref{eq.Hillas_NMM}. 
In this case, we define $a_{\nu, \rm Hillas}$ as the $\nu$MM which could give $P^{\rm HF}=\sin^21$ at some point. 
In \cref{fig.Hillas}, the dashed and dotted lines at $R>10^8$ km show such scenario, calculated via numerical evolution. 
The lines saturate to the value where $L_{\rm \nu MM}\sim L_{\rm vac}$ since, otherwise, vacuum oscillation would suppress $P^{\rm HF}$ too much for $\nu$MM to induce an $\mathcal{O}(1)$ effect.
Furthermore, the fact that the dashed lines overlap with the solid lines at $R\ll 10^9$ km, confirms that our analytic expression from \cref{sec.2.2} agrees with the numerical result, as long as the decoupling condition is satisfied.}
The benchmark values of $B=10^5$ G and $R=10^5$ km yield $a_{\nu, \rm Hillas}$ of $\mathcal{O}(10^{-12}) \mu_B$ and such values of $\nu$MM are presently unconstrained. 
These values of $B$ and $R$ are also indicated in \cref{fig.Hillas} (red triangle). 
\modify{The yellow contour lines are plotted by a joint consideration of the Hillas acceleration budget \eqref{eq.Hillas} and the proton cooling effect [case (i) in \cref{sec.3.2}]. In particular, the cooling from synchrotron radiation (adiabatic expansion) dominates at larger (smaller) $B$, resulting in a horizontal behavior (upward shift of the oblique line) \cite{Fiorillo:2021hty,Winter:2011jr}.} In Fig.~\ref{fig.Hillas}
we also show regions in $R$ and $B$ populated by some known astrophysical objects and those that fit well with the parameter space discussed above.  

In what follows, we will choose a specific astrophysical environment -- slowly rotating magnetar -- and perform a detailed numerical analysis, going beyond analytical estimates in \cref{eq.Hillas_NMM}. Magnetars yield extremely high magnetic field strength ($10^{15-17}$ G) near the surface of the star and in order to avoid significant cooling effects we will consider neutrinos generated $10^4-10^6$ km away from the star. We will also scrutinize a realization in which such neutrinos travel towards the star and enter regions with rather large magnetic field, potentially inducing a strong $\nu$MM effect.

%%%%%%%%%%%%%%%%%%%%%%%%%%%%%%%%%%%%%%%%%%%%%%%%%%%%%%%%%%%%%%%%%%%%%%%%%%%%%%%%
\section{Magnetar Systems}
\label{sec.mag}
\subsection{Magnetic Field Structure}
We approximate the neutron star spacetime by the following line element in the rest-frame of the star
\begin{align}\label{eq:equi}
    ds^2=-e^{-2\Psi}dt^2+e^{2\lambda}dr^2+r^2d\theta^2+r^2\sin^{2}\theta d\varphi^2 \,.
\end{align}
Here, $(t, r, \theta, \varphi)$ represent the Schwarzschild coordinates, and $\Psi$ is the lapse function of $r$ and $\lambda$ that is connected to the magnetar mass, $\Ms$, via
\begin{equation}
	e^{-2 \lambda} = 1 - \frac {2 \Ms} {r}\,.
\end{equation}
Note that, in \cref{eq:equi}, we have neglected the influence of magnetic field in the stellar shape and the influence of the star's spin; this allows us to perform a semi-analytical calculation.

\begin{figure}
	\centering
	\includegraphics[scale=0.39]{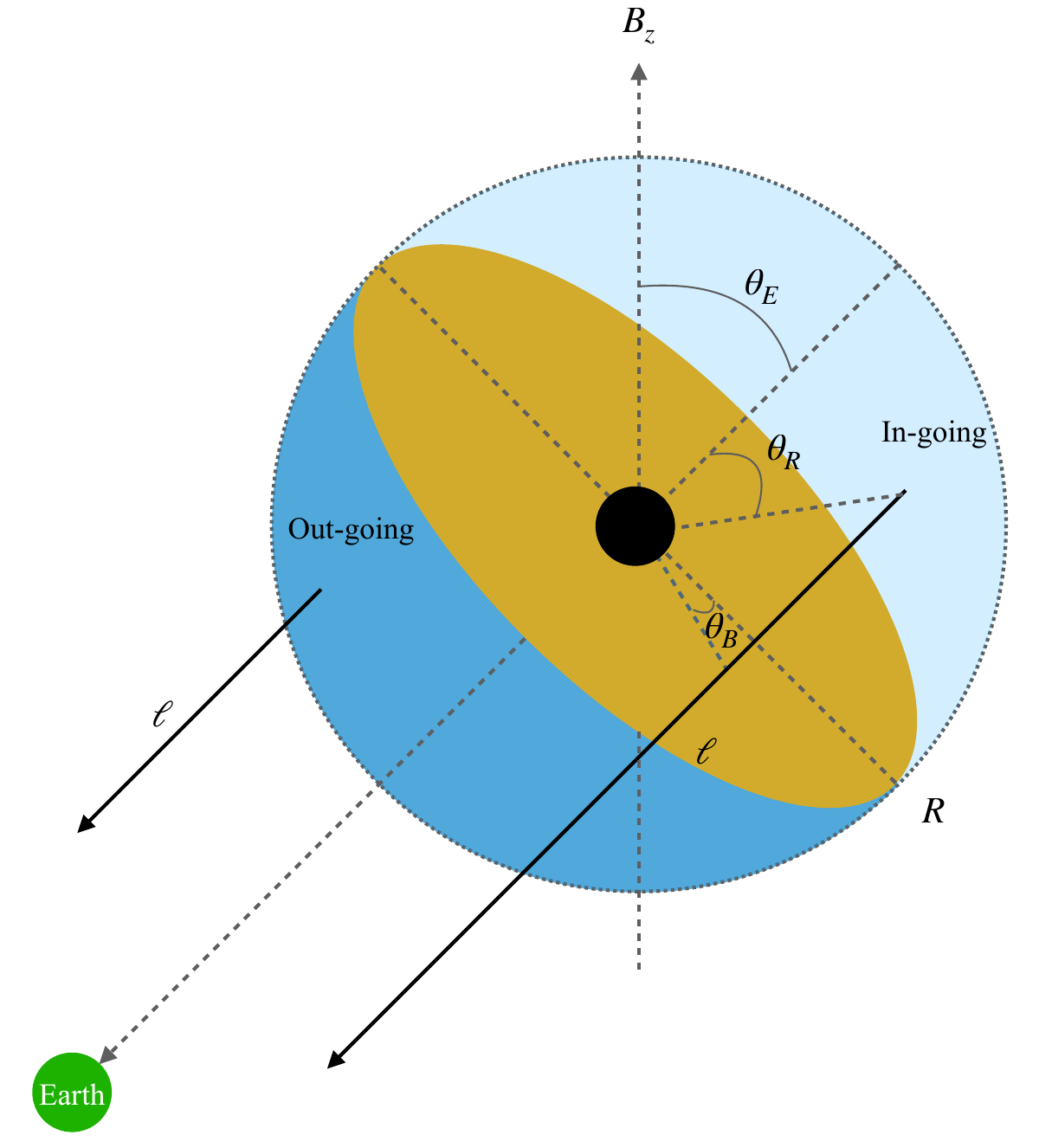}
	\caption{Sketch showing neutrino propagation from a single source. Neutrinos are produced on the shell at the distance $R$ above the stellar surface and this can be described by $\theta_B$ and $R_i=R \sin\theta_R$ coordinates for both in-going and out-going cases. The angle between the magnetic field axis and the Earth direction is $\theta_E$.}
    \label{fig.cartoon}
 \end{figure}

For simplicity, we employ dipolar magnetic field given that such configuration has been used to model the fueling of GRB afterglow for a post-merger magnetar (see e.g. \cite{DallOsso:2010uxj}). In this work, we also ignore the azimuthal (toroidal) component of the magnetic field since such structure is highly uncertain and its treatment is rather involved. By denoting the unit vector normal to a space-like hypersphere as $\hat{n}$ and by setting the magnetic axis along $z$-direction as shown in Fig.~\ref{fig.cartoon}, the dipolar magnetic field sourced by a neutron star can be expressed as
\cite{Sotani:2007be,Kuan:2021jmk}
\begin{equation} \label{eq:magf}
B^a \equiv \frac{1}{2} \hat{n}_c \epsilon^{cabd}F_{bd} = \Bc \left( 0, \frac {e^{-\lambda}} {r^2 \sin\theta} \frac{\partial\psi_2}{\partial\theta}, 
- \frac {e^{-\lambda}} {r^2 \sin\theta}  \frac{\partial\psi_2}{\partial r}, 
- \frac {\zeta(\psi_2) \psi_2 e^{-\Psi}} {r^2 \sin^2\theta}  \right).
\end{equation}
Here, $\Bc$ is the magnetic strength at the magnetar's pole and vanishing toroidal component of the magnetic field is achieved by setting $\zeta(\psi_2) = 0$. We take 
$\Bc=3\times 10^{16}$ G which could yield $B = 10^5$ G at $R=10^5$ km.

 The stream function $\psi_2$ for dipolar field has the form \cite{Mastrano:2013jaa,Suvorov:2019rzz}
\begin{align}
    \psi_2(r, \theta) = f(r)Y'_{20}(\theta)\sin\theta\,,
\end{align}
with $f(r)$ being a function related to certain boundary conditions, and $Y_{20}$ being the spherical harmonic function of degree 2 and order 0. The exterior part is characterized by
\begin{equation}\hspace*{-.2cm}
\psi_2(r,\theta) = \frac{3\Rs^3}{8\Ms^3}\left[ r^2\ln\left(\frac{r}{r-2\Ms}\right)-2\Ms r-2\Ms^2 \right]\sin^2\theta\,,
\end{equation}
with $\Rs$ being the radius of the remnant neutron star which is set as $15$~km here. This form admits that the magnetic field is force-free outside of the star and has zero-current on the stellar surface. Although the expression represents a static field, we emphasize that we are considering the slow-rotation limit of the magnetic configuration, and thus this expression is valid only within the light cylinder of neutron stars.

%%%%%%%%%%%%%%%%%%%%%%%%%%%%%%%%%%%%%%%%%%%%%%%%%%%%%%%%%%%%%%%%%%%%%%%%%%%%%%%%%
%%%%%%%%%%%%%%%%%%%%%%%%%%%%%%%%%%%%%%%%%%%%%%%%%%%%%%%%%%%%%%%%%%%%%%%%%%%%%%%%%%%%%%% 

\subsection{Simulation framework} \label{sec.simulation}
As illustrated in \cref{fig.cartoon}, after being produced at a given location on the shell with radius $R$, neutrinos propagate towards the Earth along $\vec{\ell}$ at an angle $\theta_E$ relative to the axis of the magnetic field. Although there are other directions as well, we are only interested in the portion of the flux that goes towards us, i.e.\ neutrinos streaming in a direction perpendicular to the yellow plane determined by $\theta_E$. Neutrinos produced on the upper shell (light blue) in \cref{fig.cartoon} will be in-going and those produced on the lower shell (dark blue) will be out-going with respect to the star. 
Furthermore, the production site can be projected along the direction of $\vec{\ell}$ onto the yellow plane. The projected point on the plane is determined by the distance $R_i=R \sin \theta_R$ to the stellar center and $\theta_B$, see \cref{fig.cartoon}. To summarize, for a certain magnetic field structure, the geometric parameters that enter into the computation of $\theta_\nu$ in Eq.~\eqref{eq.theta_nu} are $\theta_E$, $\theta_B$, $\theta_R$ and $R$. The two factors determining $B_\bot(\ell)$ are the distance $R_i$ and the structure of the magnetic field at $R_i$. The former is determined by $R$ and $\theta_R$ and the latter depends on $\theta_B$ and $\theta_E$.  

In our Monte Carlo simulation, these four parameters -- $\theta_E$, $\theta_B$, $\theta_R$ and $R$ -- are sampled for the in-going and out-going neutrinos, depending on the system of interest. Let us take a single source with neutrinos being produced by spherical shock wave collision. In this scenario, neutrinos are expected to be produced uniformly on the shell and emitted isotropically (see Refs. \cite{thor81,Shibata:2011kx}). This implies that at fixed $\theta_E$ and $R$, $\theta_B$ and $\theta_R$ are uniformly sampled between [0, 2$\pi$] and [$0, \pi$], respectively. Here, $\theta_R\in[0, \pi/2]$ is for in-going neutrinos and $\theta_R\in[\pi/2,\pi]$ for out-going neutrinos. The sampling of $R$ and $\theta_E$ is more involved since it depends on specific magnetar systems, namely those listed in \cref{sec.magnetar}. Nevertherless, owing to the symmetry of the magnetic field with respect to the pole and its equatorial plane, we can sample $\theta_E$ in the range $\left[0, \pi/2\right]$. To capture essential features, we will consider the following four benchmark cases with two extreme ones for the ratio of in-going and out-going neutrinos: $(i)$ in-going~:~out-going $= 1$~:~$1$ ($0<\theta_R<\pi$) and $(ii)$ in-going~:~out-going $= 0$~:~$1$ ($\theta_R>\pi/2$). As for $R$, we will simulate two cases: one with a fixed value $R=10^5$~km (the red triangle in Fig.~\ref{fig.Hillas}) and another one where this parameter is uniformly sampled within $\text{log}_{10}(R/{\rm km}) = \left[4, 6\right]$.

In \cref{sec.2.2} we have shown that the evolution operation for density matrix has a simple form, see \cref{eq.evo.oper}, provided the adiabatic condition in \cref{eq.adiab} is satisfied. 
Taking $a_\nu \sim a_{\nu,\rm Hillas}$ and $R=10^5$ km, we numerically test the level of adiabaticity by having $10\times 10 \times 10$ samples of $\theta_E$, $\theta_B$ and $\theta_R$, each over 1000 steps of $\ell \leq 10^6$ km. Among these samples and steps, we find that only $\mathcal{O}(1 \%)$ of them are not adiabatic ($\dot{\phi}/\phi \geq 0.1 a_\nu B_\bot$).

\subsection{Numerical Results}
\begin{figure}[t!]
	\centering
	\includegraphics[width=\columnwidth]{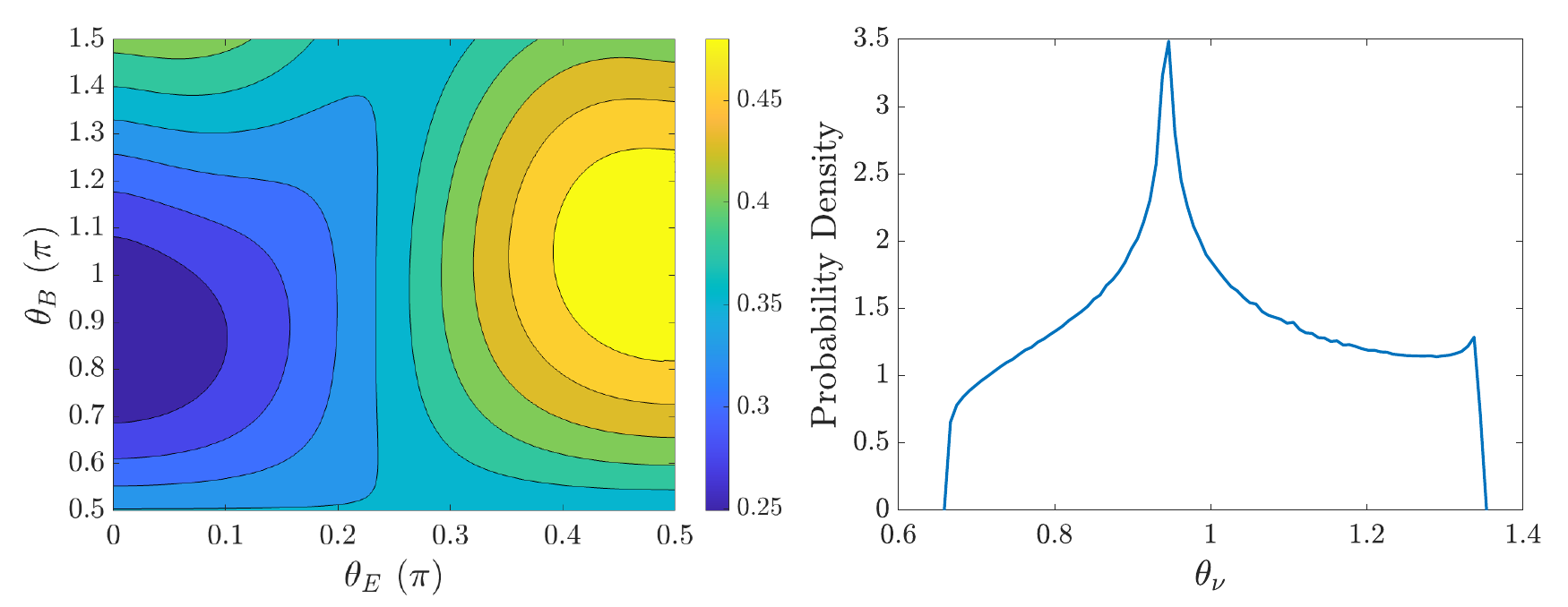}
	\caption{The left panel shows $\int d\ell \, B_{\bot}/(BR)$ as a function of $\theta_E$ and $\theta_B$. The right panel shows the probability distribution function of $\theta_\nu$ for out-going neutrinos when $a_\nu = a_{\nu,\rm Hillas}=9.1 \times 10^{-12} \mu_B$. For both panels, we have set $R=10^{5}$~km.
 }
    \label{fig.scan_th_E_B}
\end{figure}

\begin{figure}[t!]
    \centering
    \includegraphics[width=0.7\columnwidth]{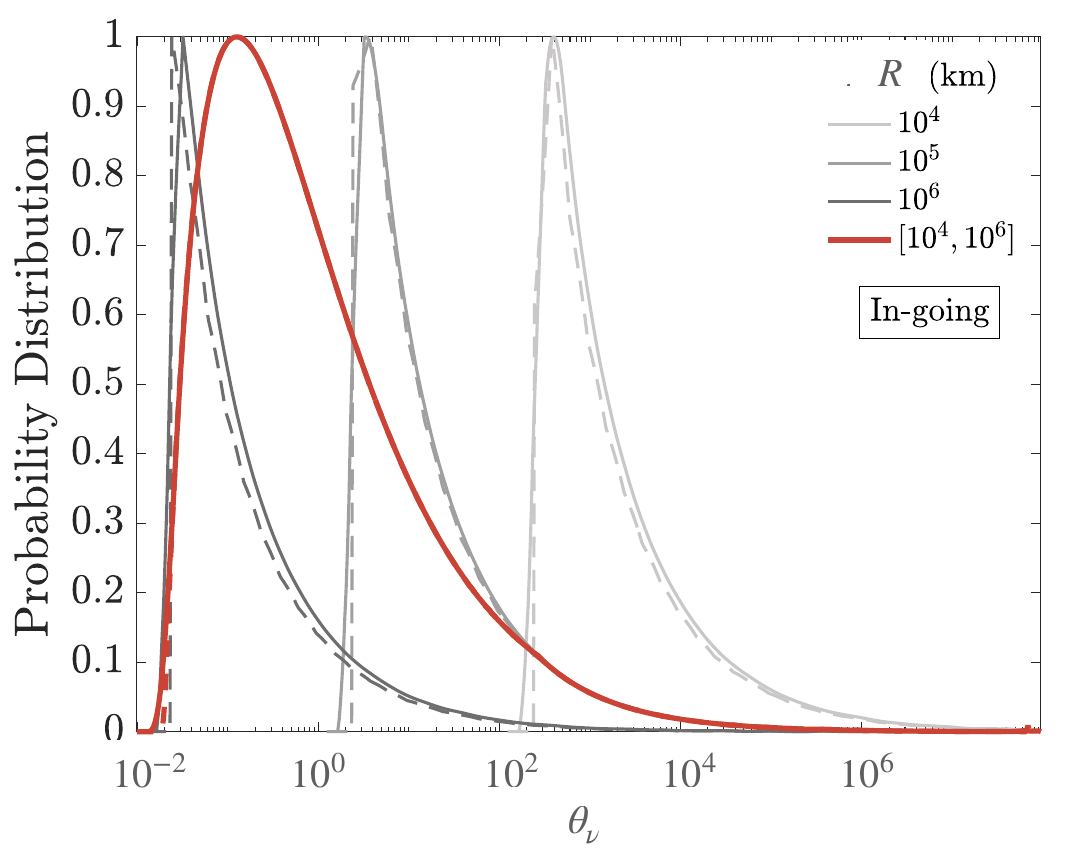}
    \caption{Normalized distribution of in-going neutrinos with $a_\nu = a_{\nu,\rm Hillas}=9.1 \times 10^{-12} \mu_B$. The three gray lines represent the case where $R$ is fixed to a certain value, and the red curve is obtained by logarithmically sampling $R$ uniformly within the region between $10^4$ and $10^6$ km. Solid gray lines represent the case where $\theta_E$ and $\theta_B$ are fixed to the values of $\pi$ and $\pi/4$, respectively. The dashed lines show the case in which these parameters are scanned over.}
    \label{fig.R_in}
\end{figure}

In this section, we explicitly obtain the probability distributions of $\theta_\nu$  using standard Monte Carlo methods for the aforementioned four benchmark cases.
We first fix $R=10^{5}$~km and obtain predictions for $\int d\ell B_{\bot}(\ell;\theta_E,\theta_B)/(BR)$ as a function of $\theta_E$ and $\theta_B$, for $\theta_R=\pi/2$ (where the in-going and out-going scenarios overlap). The result is shown in the left panel of \cref{fig.scan_th_E_B}, and we obtain the expected value of $\xi$ in \cref{eq.Hillas_NMM} to be $2.66$.
The probability distribution of $\theta_\nu$ for out-going neutrinos is obtained by scanning over $\theta_R\in [\pi/2,\pi]$, as shown in the right panel of \cref{fig.scan_th_E_B} for a fixed $a_\nu=a_{\nu,\rm Hillas} = 9.1\times 10^{-12}\mu_B$. 
On the other hand, for the in-going case, we have identified the dependence on $R_i = R\sin\theta_R$ (for $R_i\gtrsim 100$ km) as
\begin{align} \label{eq.fit_Ri}
    \theta_{\nu}(\theta_E,\theta_B;R_i)
    \approx \theta_{\nu}(\theta_E,\theta_B;,R_i=R)
    \left(
    \frac{R_i}{R}\right)^{-2}\,
\end{align}
by varying $\theta_R$ within the range $[0, \pi/2]$.
This $R_i^{-2}$ dependence can be understood since magnetic field falls as $R_i^{-3}$, and one power of $R_i$ is compensated by $\int d\ell$.
The probability distribution for the in-going neutrino case is shown in \cref{fig.R_in}, where we considered three cases with fixed $R$ (the gray curves) and one case with $R$ sampled in the range between between $10^4$ and $10^6$ km (red).

%%%%%%%%%%%%%%%%%%%%%%%%%%%%%%%%%%%%%%%%%%%%%%%%%%%%%%%%%%%%%%%%%%%%%%%%%%%%%%%%%%%%%%%
%%%%%%%%%%%%%%%%%%%%%%%%%%%%%%%%%%%%%%%%%%%%%%%%%%%%%%%%%%%%%%%%%%%%%%%%%%%%%%%%%%%%%%%

%%%%%%%%%%%%%%%%%%%%%%%%%%%%%%%%%%%%%%%%%%%%%%%%%%%%%%%%%%%%%%%%%%%%%%%%%%%%%%%%%%%%%%%
%%%%%%%%%%%%%%%%%%%%%%%%%%%%%%%%%%%%%%%%%%%%%%%%%%%%%%%%%%%%%%%%%%%%%%%%%%%%%%%%%%%%%%%

\begin{figure}
    \centering
    \includegraphics[width=0.9\columnwidth]{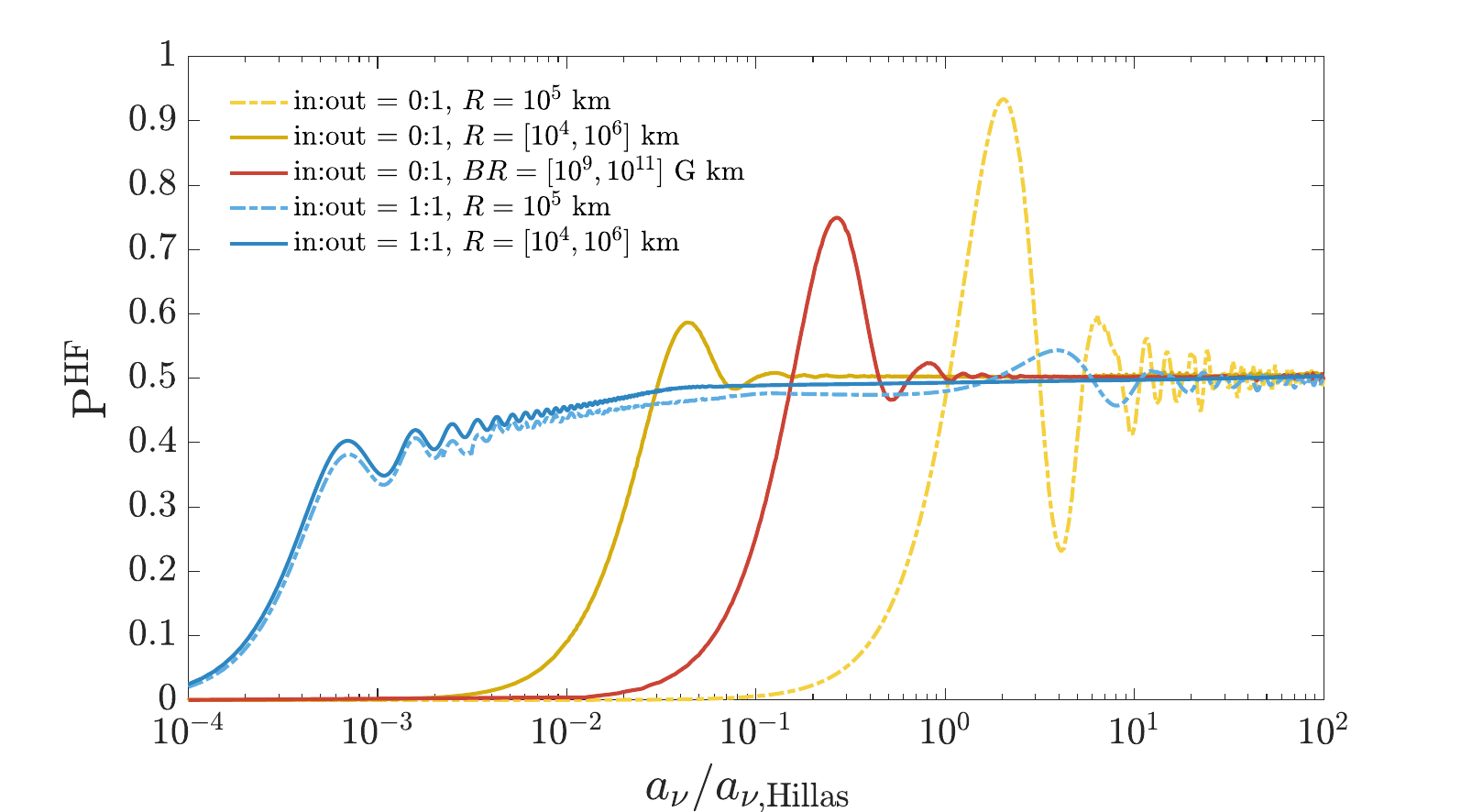}
\caption{Flavor transition HF probability (applies for transitions between any two flavors). We show cases where relevant fluxes consist only of out-going neutrinos and also scenarios in which the ratio between in-going and out-going neutrinos is $1:1$.}
    \label{fig.P_HF}
\end{figure}

Armed with $\theta_\nu$ probability distributions for both in-going and out-going case, we combine them by weighting them according to the considered flux ratio. This gives us the prediction for $\theta_\nu$, with which we can obtain HF probability, $P^{\rm HF}$. 
While in \cref{fig.scan_th_E_B,fig.R_in} we employed $a_\nu = a_{\nu,\rm Hillas}=9.1 \times 10^{-12} \mu_B$, in general one should consider a spectrum of unconstrained values for $a_\nu$. The result of such analysis is shown in \cref{fig.P_HF} where we show HF transition probability as a function of $a_\nu$. The figure shows the averaged probability; at small $a_\nu$, the contributions with largest $\int d\ell B_{\bot}$ are most relevant (the portion to the left of the peak in \cref{fig.scan_th_E_B,fig.R_in}). In this regime,  HF probability scales with $a_\nu^2$, following the behavior in \cref{eq:small} from small $\theta_\nu$ expansion. For larger values of $a_\nu$ we observe that the oscillation features develop. The prerequisite for such behavior is that probability distribution peaks around a particular value of $\theta_\nu \gtrsim 1$. Finally, when $a_\nu$ is large enough, the probability averages to $1/2$. Note that, although \cref{eq.evo.oper} may no longer apply in this scenario, the transition probabilities will still average to $1/2$.

From \cref{fig.P_HF} we can generally infer that appreciable HF transition probability occurs at $a_\nu$ values as low as approximately $a_{\nu, \rm Hillas}$ and $10^{-2}\times a_{\nu, \rm Hillas}$ for fixing $R = 10^5$ km and varying $R$, respectively, when the total relevant flux consists of out-going neutrinos. The reach in $a_\nu$ improves by $2-4$ orders of magnitude for the optimistic case where in-going and out-going neutrinos have the same contribution in the relevant flux. Additionally, to accommodate variations of different astrophysical objects, we also show, as red line in \cref{fig.P_HF}, the case where we sample over the value of $B \times R \in [10^9, 10^{11}]$ G km which is within the GRB region of \cref{fig.Hillas}. 
In summary, given that $a_{\nu , \rm Hillas}= 9.1 \times 10^{-12}\mu_B$ is our reference value, we observe that appreciable $P^{\text{HF}}$ can be obtained for $\nu$MM values as low as $\mathcal{O}(10^{-13})$ $\mu_B$ for the $0:1$ case and $\mathcal{O}(10^{-15})$ $\mu_B$ for the $1:1$ scenario. Such values are presently unconstrained and in what follows we will discuss potential signatures that they may induce at neutrino telescopes.

%%%%%%%%%%%%%%%%%%%%%%%%%%%%%%%%%%%%%%%%%%%%%%%%%%%%%%%%%%%%%%%%%%%%%%%%%%%%%%%%%%%%%%%
%%%%%%%%%%%%%%%%%%%%%%%%%%%%%%%%%%%%%%%%%%%%%%%%%%%%%%%%%%%%%%%%%%%%%%%%%%%%%%%%%%%%%%%

%=============================================================================
\section{Signatures at Neutrino Telescopes}
\label{sec:results}
The IceCube collaboration has measured flavor composition of high-energy neutrinos 
\cite{IceCube:2015gsk,HESEflavor} and has also detected a Glashow resonance event \cite{IceCube:2021rpz} that corresponds to the scattering of 6.3 PeV electron antineutrino off electron with the exchanged W boson being on shell. These two observations provide us with non-trivial information about the flavor composition of high-energy neutrinos at production sites \cite{Song:2020nfh,Huang:2023yqz,Liu:2023flr}. In particular, five types of flavor composition ($\nu_e:\nu_\mu:\nu_\tau:\bar{\nu}_e:\bar{\nu}_\mu:\bar{\nu}_\tau$) at production are typically considered. First, $(1:2:0:1:2:0)$ applies to neutrinos produced through $pp$ collisions followed by pion decay and muon decay. If muons lose significant amount of energy before decaying, the flavor composition $(0:1:0:0:1:0)$ is obtained. For the case where neutrinos are produced through $p\gamma$ collisions, followed by the decay of positively charged pion and muon, we have the flavor composition $(1:1:0:0:1:0)$, and $(0:1:0:0:0:0)$ if muon cooling is efficient. Finally, the flavor composition $(0:0:0:1:0:0)$ corresponds to neutrinos produced via neutron decay.

%%%%%%%%%%%%%%%%%%%%%%%%%%%%%%%%%%%%%%%%%%%%%%%%%%%%%%%%%%%%%%%%%%%%%%%%%%%%%%%%%
\begin{figure}
	\centering
	\includegraphics[width=0.85\columnwidth]{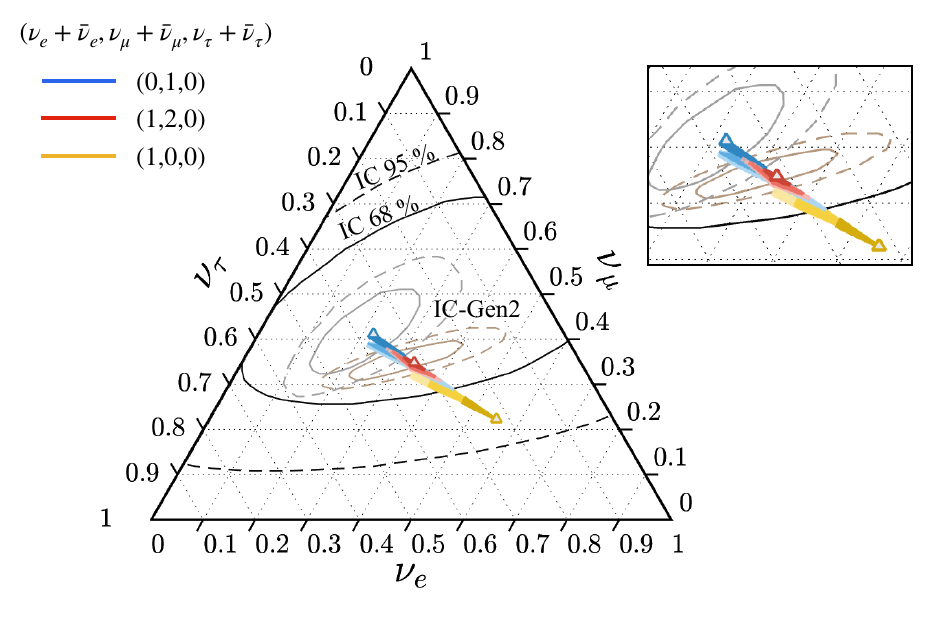}
	\caption{Ternary diagram representing flavor composition of high-energy neutrinos. The empty triangles represent the expected flavor compositions at Earth in the SM for the given initial flavor compositions. The BSM predictions for $\nu$MM are shown in blue, red and yellow for three considered flavor compositions. In producing those, all entries of the $\nu$MM matrix were considered. Relating to Fig.~\ref{fig.P_HF}, the gradient (from darker to lighter) for each color corresponds to HF probability in the range $P_{HF}\in [0,0.2], [0.2, 0.5], [0.5, 0.7]$, respectively.}
    \label{fig.FlavTri}
\end{figure}
%%%%%%%%%%%%%%%%%%%%%%%%%%%%%%%%%%%%%%%%%%%%%%%%%%%%%%%%%%%%%%%%%%%%%%%%%%%%%%%%%

%%%%%%%%%%%%%%%%%%%%%%%%%%%%%%%%%%%%%%%%%%%%%%%%%%%%%%%%%%%%%%%%%%%%%%%%%%%%%%%%% 
 \begin{figure}
	\centering
	\includegraphics[width=0.99\columnwidth]{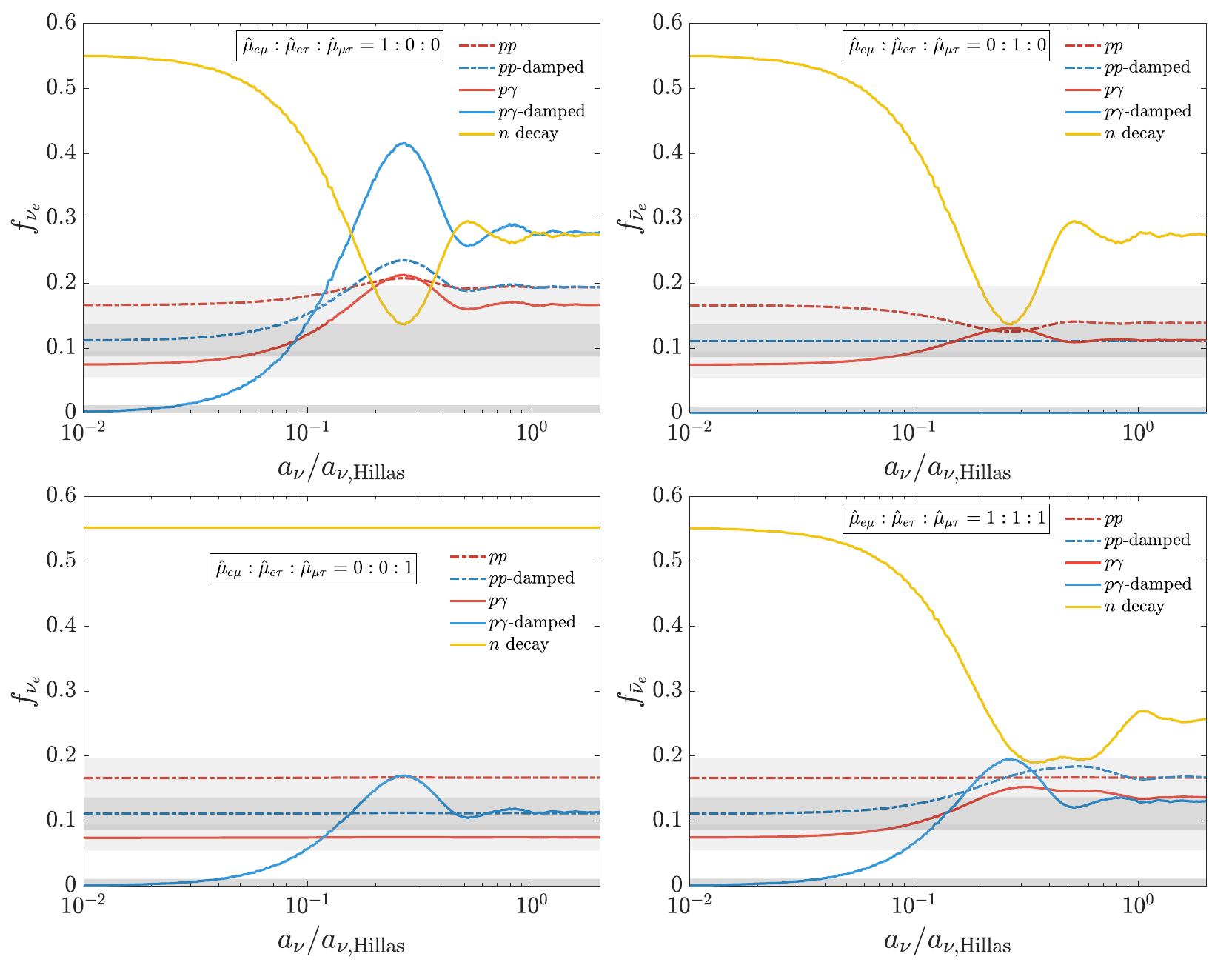}
	\caption{The fraction of the high-energy electron antineutrino flux at Earth for various production mechanisms where $BR$ is uniformly sampled within the GRB region in \cref{fig.Hillas}, and only out-going neutrinos are considered (corresponding to the red line in \cref{fig.P_HF}). Panels (a)--(d) correspond to different structures of the $\nu$MM matrix.}
    \label{fig.Glashow}
\end{figure}
%%%%%%%%%%%%%%%%%%%%%%%%%%%%%%%%%%%%%%%%%%%%%%%%%%%%%%%%%%%%%%%%%%%%%%%%%%%%%%%%% 

Since IceCube cannot distinguish between neutrinos and antineutrinos at high energies, we have instead three flavor compositions considering $(\nu_e+\bar{\nu}_e:\nu_\mu+\bar{\nu}_\mu:\nu_\tau+\bar{\nu}_\tau)$, see characteristic ternary diagram in \cref{fig.FlavTri}. The labels (empty triangles) in the plot represent the expected flavor composition in the standard case without the $\nu$MM effect and are obtained by taking present best fit values of neutrino mixing angles \cite{Esteban:2020cvm}. We show 68\% and 95\% CL limits from IceCube (black solid and dashed) and also present expected projections for measurement of flavor ratios with IceCube-Gen2 \cite{IceCube-Gen2:2020qha} (lighter gray one corresponds to $(0:1:0)$ and darker brown one represents $(1:2:0)$ at production). In what follows, we assume that future observations of high-energy neutrinos will feature plenty of events (such that robust flavor composition analysis can be performed for such subset) produced in regions with high magnetic field. Among such subset, an identification of sources would also be required for the analysis.
 
With the $P^{\text{HF}}$ from \cref{sec:results}, we propagate neutrinos from the production site to Earth and, for the three considered flavor compositions, we obtain predictions at IceCube -- see \cref{fig.FlavTri}. The obtained region corresponding to $(1:2:0)$ (red) occupies, as expected, rather narrow portion of the flavor triangle. Results are more promising for $(0:1:0)$ case; we notice that the respective region (blue) spans over significant portion of the triangle. In particular, it populates region in which flavor ratio measurements for pion decay $(1:2:0)$ sources are expected. That could cause a degeneracy between new physics and astrophysical parameters. Nevertheless, as the blue region extends even beyond the $(1:2:0)$ region, discovery of the $\nu$MM signatures is in principle possible. Another scenario (yellow) that we consider is neutron decay $(1:0:0)$. 
One can infer from the figure that such realization is in tension with the IceCube measurements at 68\% C.L. (compare yellow empty triangle with the limit). However, we observe that if new physics in the form of $\nu$MM is at play, neutron decay production mechanism becomes more consistent in light of present data since significant portion of the yellow region falls within the limit of IceCube at 68\% C.L.. We should nevertheless point out that, neutron decay production mechanism has been widely taken as less likely than the other two considered above. 

Now we turn our attention to the Glashow resonance. As stated above, this type of events can only be induced by electron antineutrinos and, in that sense, such event topology can lead to discriminating between neutrinos and antineutrinos. That appears 
especially important in light of HF that occurs due to $\nu$MM. Indeed, as we show below, the neutrino-antineutrino transitions have an intriguing interplay with certain production mechanisms. 

In \cref{fig.Glashow}, we show the fraction of high-energy electron antineutrino flux ($f_{\bar{\nu}_e}$) as a function of $a_\nu$ for several astrophysical production mechanisms. In calculating that, we employed HF transition probability which matches the one for obtaining red line in \cref{fig.P_HF}; specifically, only out-going neutrinos are considered. In \cref{fig.Glashow}, we also show bands indicating future projections at neutrino telescopes, taken from Ref. \cite{Liu:2023lxz}. Notice that, for $p\gamma$-damped scenario, $f_{\bar{\nu}_e}\to 0$ when $a_\nu\to 0$ and this is evident from all panels in which results for different textures of $\nu$MM matrix are presented. This is simply because, as discussed above, such $(0:1:0:0:0:0)$ scenario does not yield antineutrinos and hence does not lead to Glashow events. However, in the considered BSM realization, due to HF transitions, antineutrinos could get produced even for such production mechanism. This is clearly shown in \cref{fig.Glashow} for $a_\nu\neq 0$. If, in the future, at least one Glashow event from magnetar sources will be detected, $p\gamma$-damped production mechanism in such astrophysical environments may not be excluded within the $\nu$MM framework. Since we are uncertain about the origin of the detected Glashow resonance event \cite{IceCube:2021rpz}, this may be relevant already for the present data. If, on the other hand, no Glashow events will be seen from magnetar sources, it will be possible to constrain $\nu$MM, as a significant fraction of electron antineutrinos will be produced for $a_\nu \sim a_{\nu, \rm Hillas}$ across all considered production mechanisms, seen from \cref{fig.Glashow}.
%%%%%%%%%%%%%%%%%%%%%%%%%%%%%%%%%%%%%%%%%%%%%%%%%%%%%%%%%%%%%%%%%%%%%%%%%%%%%%%%
\section{Summary}
\label{sec:conclusion}
In conclusion, our work delved into a scenario where high-energy neutrinos are generated in regions characterized by intense magnetic fields. The presence of nonvanishing magnetic moments and the resulting interactions with the background magnetic field enable such neutrinos to undergo both flavor and helicity transitions. We have performed rigorous simulations of neutrino propagation and have identified potential avenues for testing magnetic moment effects with forthcoming high-energy neutrino data. For Majorana neutrinos, such effect can washout information of the production state, by averaging out the fractions of different helicities and flavors.
Specifically, for neutrino magnetic moments of $\mathcal{O}(10^{-12})$ $\mu_B$, the helicity-flipping probability could reach $\mathcal{O}(1)$ values for all considered cases and we found that this is testable in future analyses of high-energy neutrino flavor composition and Glashow events. In the optimal scenario, where high-energy neutrinos are produced at a certain distance away from the neutron star and propagate toward it, neutrino magnetic moments can be as small as $\mathcal{O}(10^{-15})$ $\mu_B$ for achieving similar effects. Given that, we have demonstrated that presently unexplored values of neutrino magnetic moment could manifest in future neutrino telescope data. Hence, this work provides a framework for experimental tests of magnetic moment-related phenomena at neutrino telescopes in the realm of high-energy neutrino interactions. 

%%%%%%%%%%%%%%%%%%%%%%%%%%%%%%%%%%%%%%%%%%%%%%%%%%%%%%%%%%%%%%%%%%%%%%%%%%%%%%%%%%%%%%

\section*{Acknowledgements}
We would like to thank Mauricio Bustamante, Kohta Murase, Guo-yuan Huang, Alexei Smirnov and Meng-Ru Wu for useful discussions.

\appendix
\section{Dirac Neutrinos}
\label{sec:dirac}
For Dirac neutrinos, the helicity basis is doubled with respect to the Majorana case, namely ~$\{h,h'\}= \{1 \,(\text{for } \nu_L), 2\,(\text{for } \bar{\nu}_L), 3 \,(\text{for } \nu_R), 4\,(\text{for } \bar{\nu}_R)\}$. The Hamiltonian defining the evolution is thus extended to a $12 \times 12$ matrix, whose non-vanishing entries are
\begin{align}
	\mathcal{H}_{\ab}^{11/22}  &= \frac{1}{2\vp}\sum_{j}U_{\alpha j}m_j^2U^{\dagger}_{j\beta} 
	\pm \delta_{\ab}V_{\alpha},  \label{eq.H_Dirac1}\\
	\mathcal{H}_{\ab}^{33}  &= \mathcal{H}_{\alpha\beta}^{44}
	= \frac{1}{2\vp}\sum_{j}U_{\alpha j}m_jU^{\dagger}_{j\beta}\,, 
	\\ \label{eq.H_Dirac2}	
	\mathcal{H}_{\ab}^{14}&= \mathcal{H}_{\ab}^{41} 
	= \left(\mathcal{H}_{\beta\alpha}^{23} \right)^{*} = \left(\mathcal{H}_{\beta\alpha}^{32} \right)^{*}
	= \mu^D_{\ab}\,B_\bot\,e^{i\phi}.
\end{align}
Here, $\mu^D$ is the $\nu$MM matrix for the Dirac neutrino case. This expression can be reduced to two $6\times 6$ matrices since $\{h,h'\}=\{1,4\}$ can be decoupled (block-diagonalized) from $\{h,h'\}=\{2,3\}$ (see the detailed treatment in \cite{Adhikary:2022phm,Kopp:2022cug}). Nonetheless, the complete $12\times 12$ expression clearly shows the difference between Dirac and Majorona neutrinos that stems from the spinor's degrees of freedom. 

The time evolution operator reads
\begin{align}\label{eq.time_opt_D}
    e^{-iA(\delta t)}=    
    \begin{pmatrix}
        \tU^D
        & 
        0
        \\     
        0
        &
        \tU^{D\dagger}
    \end{pmatrix}
    \begin{pmatrix}
        \cos\hat{\theta}^D(\delta t)
        & 
        i\sin\hat{\theta}^{D\dagger}(\delta t) 
        \\     
        i \sin\hat{\theta}^D(\delta t) 
        &
        \cos\hat{\theta}^{D\dagger}(\delta t)
    \end{pmatrix}
    \begin{pmatrix}
        \tU^D
        & 
        0
        \\     
        0
        &
        \tU^{D\dagger}
    \end{pmatrix}\,.
\end{align}
Here, the eigenvalue components are $\hat{\theta}^D\propto \diag(\theta^D_{e},\theta^D_{\mu},\theta^D_{\tau})$.

The flavor structure for Dirac neutrinos is represented by $\tU^D$ which diagonalizes $\nu$MM matrix $\mu^D$. If $\mu_D$ is already diagonal in flavor space, then $\diag(\theta^D_{e},\theta^D_{\mu},\theta^D_{\tau})=\diag(\mu^D_{ee},\mu^D_{\mu\mu},\mu^D_{\tau\tau})$, and $\tU^D$ is unitary. In fact, in this case, the Hamiltonian can be reduced to three $2\times2$ matrices, one for each flavor. 

\section{Candidates for the magnetar system} \label{sec.magnetar}
We enumerate scenarios for high-energy neutrino production in highly magnetized astrophysical environments, focusing on magnetars as emitters of jet and/or central engines for GRBs.

\begin{enumerate}[label=$(\roman*)$]
%%%%%%%%%%%%%%%%%%%%%%%%%%%%%%%%%%%%%%%%%%%%%%%%%%%%%%%%%%%%%%%%%%%%%%%%%%
\item Young magnetars tend to bear a rapid, and differential rotation, and posses a strong magnetic field with non-trivial multipolar structure. It can therefore be envisioned that the spin axis may differ from the magnetic field's axis. The unipolar induction of a rotating, magnetized neutron star will render an electric potential, which can possibly reach a magnitude of
	\begin{align}
	    \Phi_{\rm max}=\frac{\Omega^2 \Bc R_\star^3}{2c^2}\,,
	\end{align} 
    in the vicinity of the stellar surface \cite{Gold69}.
    Here, $\Omega$ denotes the magnitude of stellar spin and $R_\star$ is the radius of the neutron star. The associated electromotive force then accelerates the charged particles that constitute a plasma. For the cases where $\vec B$ and $\vec{\Omega}$ satisfy the condition $\vec{B}\cdot\vec{\Omega}<0$, the positively charged particles, including protons, will be unleashed along the open field lines in the polar regions \cite{rude75}. In particular, the injection rate of protons to the stellar wind \cite{aron03} may be approximated by the Goldreich-Julia rate \cite{carp20}.
    Depending on the surface temperature of the pulsar, the relativistic protons are expected to hit photons and generate high-energy neutrinos via $\Delta$-resonance \cite{Link05,Link06},
	\begin{align}\label{eq:DeltaRes}
	    p\gamma\rightarrow \Delta\rightarrow n\pi^{+}\rightarrow n\nu_\mu\mu^+\rightarrow n\nu_\mu e^+\nu_e\overline{\nu}_\mu\,,
	\end{align}
  It should be noted that the produced pions and muons will also undergo several cooling mechanisms and will hence only be able to hand over a fraction of the energy to neutrinos \cite{razz04}. 
%%%%%%%%%%%%%%%%%%%%%%%%%%%%%%%%%%%%%%%%%%%%%%%%%%%%%%%%%%%%%%%%%%%%%%%%%%%%%%%%

%%%%%%%%%%%%%%%%%%%%%%%%%%%%%%%%%%%%%%%%%%%%%%%%%%%%%%%%%%%%%%%%%%%%%%%%%%%%%%%%%%
\item Within fireballs \cite{pira99} that scintillate gamma-ray bursts -- both long and short ones -- protons and electrons will undergo Fermi acceleration by the magnetic field. Depending on the properties of fireballs (see \cite{He12} and references therein), some protons can be accelerated sufficiently to enable $pp$ collisions. This collision will create mesons which further decays to produce neutrino transients \cite{Waxman:1997ti,Kimura:2017kan}. These $pp$ collisions mainly lead to pion production though other mesons (e.g. kaons) can also occur \cite{Lindsey:1989ni}. The fact that pions are mostly produced, however, does not necessarily imply that they are the main source of neutrinos; the less efficient radiative cooling and the shorter lifetime of kaons arguably make kaons more important source of neutrinos for energy range above TeV (thus applies to the energy range considered in the main text).
We should also stress that the $p\gamma$ collisions are also at play and yield photomesons. Namely, in addition to the process shown in \cref{eq:DeltaRes}, kaons can be produced as well and subsequently decay to neutrinos \cite{ando05}
    \begin{align}
    	p\gamma\rightarrow \Lambda^0 K^+,\,\, \Sigma^0K^+,\,\,\Sigma^+K^0.
    \end{align}
%%%%%%%%%%%%%%%%%%%%%%%%%%%%%%%%%%%%%%%%%%%%%%%%%%%%%%%%%%%%%%%%%%%%%%%%%%%%%%%%%

%%%%%%%%%%%%%%%%%%%%%%%%%%%%%%%%%%%%%%%%%%%%%%%%%%%%%%%%%%%%%%%%%%%%%%%%%%%%%%%%%%%
	\item At early stages of millisecond magnetars, occurring following either a binary merger or a supernova, the relativistic wind spewed from the central remnant will be braked by its interaction with ejecta, producing shocks heating up the ponderable medium. Within these pulsar wind nebula, inelastic $pp$ collisions are expected to effectively operate, giving rise to a copious number of mesons \cite{Murase:2009pg}, e.g. neutral and charged pions \cite{Fang:2017tla}. These mesons may, in principle, decay and thereby generate high-energy neutrinos; yet, no such events have been detected \cite{IceCube:2020svz}.
%%%%%%%%%%%%%%%%%%%%%%%%%%%%%%%%%%%%%%%%%%%%%%%%%%%%%%%%%%%%%%%%%%%%%%%%%%%%%%%%%%%%%

%%%%%%%%%%%%%%%%%%%%%%%%%%%%%%%%%%%%%%%%%%%%%%%%%%%%%%%%%%%%%%%%%%%%%%%%%%%%%%%%%%%%%%%%%
	\item The $pp$ collisions within cocoon systems, formed atop the remnant magnetar of binary merger \cite{Kimura:2018vvz} or supernovae \cite{Murase:2013ffa}, will seed middle stage mesons such as pions and kaons. Their decay into highly energetic neutrinos may, however, be stagnated considerably since the accelerated mesons will be cooled via several mechanisms (e.g., \cite{razz04,ando05}) resulting in considerable energy loss. Owing to the longer cooling times and shorter lifetimes, there are speculations that charm contribution to the neutrino population may be more important than expected \cite{carp20}.
%%%%%%%%%%%%%%%%%%%%%%%%%%%%%%%%%%%%%%%%%%%%%%%%%%%%%%%%%%%%%%%%%%%%%%%%%%%%%%%%%%%%%%%%%%
%%%%%%%%%%%%%%%%%%%%%%%%%%%%%%%%%%%%%%%%%%%%%%%%%%%%%%%%%%%%%%%%%%%%%%%%%%%%%%%%%%%%%%%%%%%
	\item On top of the aforementioned proton-involved collisions, neutrons in the relativistic outflow may activate production of metastable mesons (e.g., \cite{Murase:2013hh}). Such events may be detected in future analyses of the brightest GRB recorded to date (GRB 221009A) \cite{mura22}.
%%%%%%%%%%%%%%%%%%%%%%%%%%%%%%%%%%%%%%%%%%%%%%%%%%%%%%%%%%%%%%%%%%%%%%%%%%%%%%%%%%%%%%%

\end{enumerate}

%-----------------------------------------------------------------------------
\bibliographystyle{JHEP}
\bibliography{refs}
%-----------------------------------------------------------------------------

\end{document}